\documentclass[letterpaper,preprint,showpacs,floatfix]{revtex4} 
\RequirePackage{graphicx} 
\newcommand{\D}{\mathrm d}
\newcommand{\I}{\mathrm i}
\newcommand{\E}{\mathrm e}
\newcommand{\Ai}{\mathrm {Ai}}
\newcommand{\Bi}{\mathrm {Bi}}
\newcommand{\curl}{\mathrm {curl}}

\begin{document} 

\title{Communication through plasma sheaths} 
 
\date{\today} 
 
\author{A.~O.~Korotkevich} 
\email{kao@itp.ac.ru} 
\affiliation{Landau Institute for Theoretical Physics RAS,\\
2, Kosygin Str., Moscow, 119334, Russian Federation} 
 
\author{A.~C.~Newell} 
\email{anewell@math.arizona.edu} 
\affiliation{Department of Mathematics, The University of Arizona,\\
617 N. Santa Rita Ave., Tucson, AZ 85721, USA}
 
\author{V.~E.~Zakharov} 
\email{zakharov@math.arizona.edu} 
\affiliation{Department of Mathematics, The University of Arizona,\\
617 N. Santa Rita Ave., Tucson, AZ 85721, USA}
\affiliation{Lebedev Physical Institute RAS,\\
53, Leninsky Prosp., GSP-1 Moscow, 119991, Russian Federation} 
\affiliation{Landau Institute for Theoretical Physics RAS,\\
2, Kosygin Str., Moscow, 119334, Russian Federation}
\affiliation{Waves and Solitons LLC, 918 W. Windsong Dr., Phoenix, AZ 85045, USA} 

\pacs{52.40.Kh, 52.35.Mw, 52.40.Db}
\begin{abstract} 
We wish to transmit messages to and from a hypersonic vehicle around which a plasma sheath has formed.
For long distance transmission, the signal carrying these
messages must be necessarily low frequency, typically 2 GHz, to which the plasma sheath is opaque. The idea
is to use the plasma properties to make the plasma sheath appear transparent. 
\end{abstract} 
\maketitle

\section{Introduction.}
\subsection{General discussion.}
A vehicle moving through the stratosphere (altitudes 40km-50km) at hypersonic velocities (8-15 Mach) is covered
by a plasma sheath. Typically, the plasma density $n$ can be as high as $10^{18}m^{-3}$ with corresponding plasma frequency
\begin{equation}
\label{plasma_freq}
2\pi f_L = \omega_L = \left(\frac{e^2 n}{M \varepsilon_0}\right)^{1/2}
\end{equation}
of about $9 GHz$. In (\ref{plasma_freq}), $e$ is the electron charge $-1.6\times 10^{-19} C$,
$\varepsilon_0 = 8.85\times 10^{-12} CV^{-1}m^{-1}$ and $M$ is the electron mass $9\times 10^{-31} kg$.
Therefore the plasma is opaque to frequencies lower than $9 GHz$. Direct communication through such a plasma
to and from the vehicle is impossible because frequencies $f$ suitable for long distance propagation through
the atmosphere are usually much less. For example, the standard frequency used for navigational satellite systems,
including the global positioning system (GPS), are less than $2 GHz$. For the GPS, $f = 1.57542 GHz$.

The challenge is to devise means to maintain continuous contact with the hypersonic vehicle. When such vehicles
were principally spacecrafts, a blackout period of up to two minutes was acceptable albeit undesirable. But
when the vehicles are of military origin, it is clear that continuous contact is essential for both targeting
and rapid abort reasons.

It is a challenge which has drawn many responses. They fall into several categories. The first ignores the presence
of the plasma by using signals with frequencies well above the plasma frequency. The difficulty with this method
is that such signals are heavily attenuated in and scattered by the atmosphere. A second means, which also ignores
the plasma, is to use low frequency signals in the $100 MHz$ range where wavelengths are large compared to the
plasma sheath thickness (typically of the order of a meter). But such solutions have high cost and low bit rates
and are not well supported by existing infrastructure. A third category of solutions violates the plasma.
One approach is to remove, by vehicle reshaping, for example, the plasma from certain points on the vehicle at which
one might place an antenna. Another is to destroy it by electrophilic injection or by injecting water drops.
A third approach is to use powerful magnets to reshape the plasma. Such solutions involve a heavy cost in that
design features necessary for their
implementation must be built into the vehicle a priori. Nevertheless some are feasible and worthy of consideration.
For example, it is possible to build an antenna into a sharp leading edge which would protrude beyond the plasma
and survive for sufficiently long (it would be eventually destroyed by ablation) to cover the flight time.

The fourth category of solutions, and the one to which we are attracted, uses the properties of the plasma itself
to effect transmission in the same way a judo expert uses the strength and motion of an opponent to defeat him.
One idea is to create new modes of oscillation and propagation by the introduction of magnetic fields. Indeed, for strong
enough fields, the Larmor frequency $f_{Larmor}$ is sufficiently large that the window $(f_{Larmor}, \max(f_L))$ for which the plasma
is opaque is small and transmission can be achieved for frequencies below $f_{Larmor}$. But the introduction of magnetic
fields involves large additional weight and new design features.
The second idea is much more simple. Its aim is to take advantages of nonlinear properties of plasma to render
it effectively transparent to the signal. Communications both to and from the vehicle are feasible using basically
the same ideas. We shall first describe the ``to the vehicle'' case. Consider Figure \ref{plasma_response_fig}
in which we show schematically the response of the plasma to an incoming signal with low frequency $\omega$
from a direction which makes an angle $\phi$ with the normal to the vehicle. There are two principal features
to the response. First, there is a reflection from the layer at a point $z=z_r$ where the plasma frequency
at the point $\omega_L (z_r)$ is $\omega\cos\phi$. However, the influence of the signal is felt beyond that
point, namely at the resonant layer $z=0$ where $\omega_L (0) = \omega$. Langmuir oscillations are excited there
which produce large transversal and longitudinal components of the electric field. The resonant layer acts
as an antenna. The task is to find a way to connect the antenna at the resonant layer at $z=0$ to a receiver
on board the vehicle at $z=R$. There are several possibilities which we have outlined before \cite{Nazarenko1994, NNR1994, NNR1995}.

The most practical one, however, is also the most simple and first suggested without a detailed numerical
simulation in \cite{Nazarenko1994}.
We use an onboard source, which we call the pump,
to generate electromagnetic signals of sufficiently high frequency $\omega_p$
($\omega_p > \max\limits_{z}^{}\omega_L (z) + \omega$) that they can propagate through the plasma. There are several
candidates for such a source. For example, available on the open market is a klystron amplifier which can generate
$3 kW$ of power at frequencies of $12-14 GHz$. These high frequency waves have only to travel distances of a meter or less.
They interact nonlinearly with and scatter off the signal wave. Not surprisingly, the largest contribution to the scattered
wave comes from the nonlinear interaction of the pump wave with the plasma density distortion induced by the incoming signal wave
at the resonant layer. We call the scattered wave a Stokes wave because the scattering process is a three wave interaction
analogous to Raman scattering. The Stokes wave with frequency $\omega_S = \omega_p - \omega$ carries the information encoded
on the signal wave back to the vehicle. We will show that, whereas much of the scattered Stokes wave propagates away from the
vehicle, a significant fraction is returned to the vehicle.

What is remarkable is this. The ratio of the power flux of the Stokes wave received at the vehicle to the power flux contained
in the signal wave at the plasma edge can be between 0.7 and 2 percent. This means that reception of GPS signals may be possible
because one simply needs an onboard receiver approximately 100 times more sensitive than commercially available hand held receivers or use sufficiently larger antenna.
We shall discuss in the conclusion the sensitivity required for a variety of sources.

Communications from the vehicle requires two power sources on the vehicle. One, which we term the Stokes wave generator, will
also carry the signal. The other is the pump wave. Both have carrier frequencies above that of the maximum of the plasma frequency.
Their nonlinear interaction in the plasma produces an oscillations of frequency $\omega = \omega_p -\omega_S$.
Consider Figure \ref{from_concept}. For $z_r < z < R$ where $z_r$ is determined by $\omega_L (z_r) = \omega\cos\phi$
and $\phi$ is calculated from the differences in propagation directions of the pump and Stokes waves, the oscillation does not
propagate and its strength decays away from the vehicle. Nevertheless this oscillation is sufficiently strong to act as a power
source for a propagating wave in the region $z < z_r$ where $\omega\cos\phi > \omega_L (z)$. In the conclusion we
analyze what power is required in order for the signal to be detected by distant receivers. It appears that
even if we use usual available on the market generators communication can be put into practice.

\subsection{Plan of the paper.}
The plan of the paper is as follows. We begin in {\it Section~2} with a detailed analysis of the two dimensional propagation and
interaction of a signal wave of frequency $\omega$, a pump wave of frequency $\omega_p$ and a Stokes wave of frequency
$\omega_S$ through a plasma with a given density profile $n_0 (z)$ where $z$ is the direction normal to the vehicle. The key
equation is a modification of the well known Ginzburg equation \cite{Ginzburg_bib}
\begin{eqnarray}
\label{Ginzburg_equation}
\frac{\partial}{\partial z}\left(\left(\frac{\varepsilon_0}{\varepsilon(z,\Omega)}\right)\frac{\partial\vec H}{\partial z}\right) +
\frac{\varepsilon_0}{\varepsilon(z,\Omega)}\frac{\partial^2\vec H}{\partial y^2} + \frac{\Omega^2}{c^2}\vec H =\\ =-\left[\nabla\times\left(\frac{\varepsilon_0}{\varepsilon(z,\Omega)}\vec j_{NL}\right)\right],\nonumber
\end{eqnarray}
for the magnetic field amplitude $(H(y,z), 0, 0)\E^{-\I\Omega t}$ of an oscillation of frequency $\Omega$.
In (\ref{Ginzburg_equation}), the effective electric susceptibility is
\begin{equation}
\label{epsilon_z_Omega}
\varepsilon (z, \Omega) = \varepsilon_0 \left( 1 - \frac{\omega_L^2 (z)}{\Omega^2}\left(\frac{1}{1+\I\nu/\Omega}\right)\right),
\end{equation}
($\omega_L (z)$ is the local plasma frequency and $\nu$ the collision frequency). The susceptibility is due to the linear response
of the plasma to the electric fields of whichever waves are involved. The nonlinear current $\vec j_{NL}$ will be determined both by
the product of the plasma density distortion with the linear current and the nonlinear response of the electric velocity field
due principally to dynamic pressure forces. We observe that, for $\Omega \gg \max\limits_{z}^{}\omega_L (z)$, the electric susceptibility
is approximately $\varepsilon_0$ and the left hand side of the nonlinear Ginzburg equation (\ref{Ginzburg_equation}) is
the usual wave operator.

How do we use (\ref{Ginzburg_equation})? For the case of communication to the vehicle, we use it in two ways. First with
$\vec j_{NL}=0$, we determine for $\Omega = \omega$ and $H(y,z) = H(z)\E^{\I(\omega/c)y\sin\phi}$, the field $H(z)$ from which
the distortion to the plasma produced by the incoming wave is calculated. In this instance, $H(z)$ satisfies
\begin{eqnarray}
\frac{d^2 H}{d z^2} -
\frac{1}{\varepsilon(z,\omega)}\frac{d \varepsilon(z,\omega)}{d z}\frac{d H}{d z} +\label{Ginzburg_to_equation}\\
\frac{\omega^2}{c^2}\left(\frac{\varepsilon(z,\omega)}{\varepsilon_0} - \sin^2 \phi\right)H = 0.\nonumber
\end{eqnarray}
A glance at the third term shows that propagation is impossible for $\varepsilon/\varepsilon_0 < \sin^2 \phi$
or, from (\ref{epsilon_z_Omega}), for $\omega\cos\phi < \omega_L (z)$. The importance of the resonance layer
where $\varepsilon (z,\omega) \simeq 0$ is seen from the denominator in the second term. Having solved for $H(z)$
from (\ref{Ginzburg_to_equation}) we can then calculate the plasma distortion field $\delta n(z)$. Its interaction
with the pumping wave then produces a nonlinear current $\vec j_{NL}$ which gives rise to the Stokes
wave. The Stokes wave $H_S (y,z)$ and its propagation is calculated by solving (\ref{Ginzburg_equation}) with this
$\vec j_{NL}$ and appropriate boundary conditions at the plasma edge and at the vehicle. Our goal is to determine
$H_S (y, z = R)$. We give the results of both the numerical simulation and an analytic estimation. The latter takes
advantage of the fact, that, for the Stokes wave, $\omega_S \gg \max\limits_{z}^{}\omega_L (z)$ and that the principal plasma
distortion occurs at the resonance layer.

For communicating from the vehicle, we solve (\ref{Ginzburg_to_equation}) with the right hand side given by
$-\nabla\times\frac{\varepsilon_0}{\varepsilon}\vec j_{NL}$ with $\vec j_{NL}$ calculated from the nonlinear interaction of the pump and Stokes waves.
Here the goal is to calculate the flux of power of the signal wave with frequency $\omega = \omega_p - \omega_S$
as it leaves the plasma edge in the direction of some distant receiver.

In {\it Section 3}, we describe the numerical procedure and give detailed results of our calculations.

Finally, in the {\it Conclusion}, we use our results to calculate the powers of both the incoming and outgoing signals
at their respective receivers. We discuss in addition several important considerations:
\begin{itemize}
\item The advantages, particularly in terms of available power, of using pulsed signals.
\item The possibility of using GPS sources for incoming signals.
\item The challenges involved in making ideas practicable.
\end{itemize}

\section{Analytics.}
\subsection{Basic theory.}
We shall study a very idealized situation when the plasma sheath
is a flat slab. The plasma density is a linear function of the
horizontal coordinate $z$
\begin{equation}
n_0 (z) = n_0 \frac{z+L}{R+L}.
\end{equation}
In this geometry the vehicle is the vertical wall placed at
$z=R$. The plasma density near the vehicle is $n_0$. The plasma contacts the vacuum at $z=-L$,
where $n=0$. We shall study two
situations: communication to the vehicle and communication
from the vehicle. In both cases, three almost monochromatic
electromagnetic waves exist in the plasma. Two of them have
high frequencies $\omega_p$ (pumping wave), $\omega_S$ (Stokes
wave). The third one has low frequency $\omega$, satisfying
the condition
\begin{equation}
\omega = \omega_p - \omega_S.
\end{equation}
In the ``to the vehicle'' case $\omega$ is the circular frequency of the incoming signal.
In the ``from the vehicle'' case, $\omega$ is the circular frequency of the outgoing signal.
In both these cases, the low-frequency signal plays a key role. Because the local plasma
frequency at $z = 0$ is $\omega$,
\begin{equation}
\omega^2 = \frac{e^2 n_0}{M \varepsilon_0}\frac{L}{R+L}.
\end{equation}
Let us denote also the Langmuir frequency at the vehicle as
$$
\omega_L^2 = \frac{e^2 n_0}{M \varepsilon_0}.
$$
Thus
$$
\frac{L}{R+L} = \frac{\omega^2}{\omega^2_L} = \frac{f^2}{f^2_L}.
$$
In a realistic situation $f_L\simeq 9 GHz$ (it corresponds to
$n_0 = 10^{18} m^{-3}$), $f \simeq 2 GHz$, $R+L = 1 m$, and
$L \simeq 0.05 m$.
The wavelength of the incoming signal in the vacuum is
$\lambda = c/f = 0.15 m$, so that $\lambda > L$. We point out
that in the case of low-frequency wave reflection from the
ionosphere, the situation is the opposite $\lambda << L$.

We shall assume that the ions' positions are fixed and the plasma is
cold ($T_e \simeq 0$). The magnetic field has only one component
$H_x$. The electric field has two components $E_y$, $E_z$.
Neither the electric nor magnetic fields depend on the
$x$-coordinate. Maxwell's equations read
$$
\vec E (0, E_y(y,z), E_z (y,z));\;\; \vec H (H(y,z), 0, 0)
$$
\begin{eqnarray}
\nabla \times \vec E = -\mu_0 \frac{\partial \vec H}{\partial t},\label{rot_E}\\
\nabla \times \vec H = \varepsilon_0 \frac{\partial \vec E}{\partial t} + \vec j,\label{rot_H}\\
\nabla \cdot \vec H = 0,\label{div_H}\\
\nabla \cdot \varepsilon_0 \vec E = e (n - n_0(z)),\; \vec j = en\vec v.\label{div_E}
\end{eqnarray}
\begin{eqnarray}
\frac{\partial \rho}{\partial t} + \nabla \cdot \vec j = 0\nonumber\\
\frac{\partial n}{\partial t} + \nabla \cdot n\vec v = 0\label{Continuity},
\end{eqnarray}
\begin{equation}
\label{Euler}
\frac{\partial \vec v}{\partial t} + \nu \vec v = \frac{e \vec E}{M} + \vec v \times\left(\left[\nabla\times\vec v \right] + \frac{\mu_0 e}{M}\vec H\right) - \frac{1}{2}\nabla v^2,
\end{equation}
\begin{eqnarray*}
c = \frac{1}{\sqrt{\varepsilon_0\mu_0}}\simeq 3\times10^{8}ms^{-1},\\
n_0 \simeq 10^{18}m^{-3},\;
\omega_{L}^2(R) = \frac{e^2n_0}{M\varepsilon_0},\\ \frac{\omega_L(R)}{2\pi} = f_L(R) = 9 GHz.
\end{eqnarray*}
The power flux in vacuum is
$$
S = 2\varepsilon_0 c \left|E\right|^2 = 2c \mu_0\left|H\right|^2 Wm^{-2};\;1 Wm^{-2} \rightarrow 13.7 Vm^{-1}.
$$
In equation (\ref{Euler}) $\nu$ is the effective friction of
the electron fluid with the neutral gas, sometimes called the ion collision frequency.
We take $\nu = 10^8 Hz$.

The current $\vec j = \vec j_L + \vec j_{NL}$. $\vec j_L$ is
the linear response of the plasma on the electric field, $\vec j_{NL}$
is the current due to nonlinear effects. For a
monochromatic wave of frequency~$\Omega$, Maxwell's equations
can be rewritten in the following form
\begin{eqnarray*}
\nabla\times\vec H &=& -\I\Omega\varepsilon\vec E + \vec j_{NL},\;\varepsilon_0 \nabla\times\vec E = \I\varepsilon_0 \mu_0 \Omega \vec H,\\
\I\frac{\Omega}{c^2} \vec H &=&
\frac{\I}{\Omega}\nabla\times\left(\frac{\varepsilon_0}{\varepsilon}\nabla\times\vec H\right) -
\frac{\I}{\Omega}\nabla\times\left(\frac{\varepsilon_0}{\varepsilon}\vec j_{NL}\right),
\end{eqnarray*}
\begin{equation}
\label{ProtoGinzburg}
\frac{\Omega^2}{c^2}\vec H = \nabla\times\left(\frac{\varepsilon_0}{\varepsilon}\nabla\times\vec H\right) - 
\nabla\times\left(\frac{\varepsilon_0}{\varepsilon}\vec j_{NL}\right).
\end{equation}
In our geometry, (\ref{ProtoGinzburg}) is one scalar equation.
We should stress that this is an exact equation. The only challenge
is the calculation of $\vec j_{NL}$.

Finally for the magnetic field, one obtain the Ginzburg equation
\begin{eqnarray}
\frac{\partial^2 H}{\partial z^2} &-& \frac{\varepsilon'}{\varepsilon}\frac{\partial H}{\partial z} +
\frac{\varepsilon}{\varepsilon_0}\frac{\Omega^2}{c^2}H+\nonumber\\
+ \frac{\partial^2 H}{\partial y^2} &=& 
- \left(\nabla\times\vec j_{NL}\right)_x - \frac{\varepsilon'}{\varepsilon} (\vec j_{NL})_y,\;
\varepsilon' = \frac{\partial \varepsilon}{\partial z}.
\label{Ginzburg}
\end{eqnarray}
For the high frequency pump and Stokes waves $\varepsilon\simeq\varepsilon_0$.
Some exact solutions of simplified versions of the homogeneous Ginzburg equation
for several important cases can be found in Appendix \ref{APP:Analytics_homogenious}.

What we are going to do is the following: in subsection \ref{Lin_resp} we shall
calculate linear responses of the plasma to an electromagnetic wave, such as the electron
velocity, linear current and the electron density profile perturbation;
the calculation of the first nonlinear correction to the linear current is done in
subsection \ref{Nonlin_curr}; analytic estimations for ``to the vehicle'' and
``from the vehicle'' cases are given in subsections \ref{Analytics_to} and
\ref{Analytics_from} respectively.

\subsection{\label{Lin_resp}Linear responses.}
In order to calculate the nonlinear current we need to consider
the linear responses of the plasma to the presence of an
electromagnetic wave.
For a field with frequency $\Omega$
$$
H \sim \E^{-\I\Omega t},
$$
from (\ref{Euler}), the linear term in the velocity
\begin{equation}
\label{Linear_v}
\vec v_L = \frac{\I e}{M \Omega}\frac{1}{1+\I\nu/\Omega}\vec E,
\end{equation}
and
$$
\vec j_L = \frac{\I e^2 n_0}{M \Omega}\frac{1}{1+\I\nu/\Omega}\vec E.
$$
From (\ref{rot_H})
$$
\nabla\times\vec H = -\I\Omega\varepsilon_0 \vec E + \frac{\I e^2 n_0}{M\Omega}\frac{1}{1 + \I\nu/\Omega}\vec E =
-\I\Omega\varepsilon\vec E
$$
Using Maxwell equations one can express all responses
in terms of magnetic field
\begin{eqnarray}
\vec E = \frac{\I}{\Omega\varepsilon(\Omega)}\left(0, \frac{\partial}{\partial z}H, -\frac{\partial}{\partial y}H\right),\\
\vec v_L = -\frac{e}{M\Omega^2\varepsilon(\Omega)}\frac{1}{1 + \I\nu/\Omega}
\left(0, \frac{\partial}{\partial z}H, -\frac{\partial}{\partial y}H\right),\label{linear_velocity}\\
\vec j_L = -\left(1 -\frac{\varepsilon_0}{\varepsilon(\Omega)}\right)
\left(0, \frac{\partial}{\partial z}H, -\frac{\partial}{\partial y}H\right).
\end{eqnarray}
The expression for a distortion $\delta n$ of the electron density in
the plasma $n (z) = n_0 (z) + \delta n (y,z,t)$
can be derived from (\ref{Continuity}) and (\ref{linear_velocity}),
\begin{equation}
\delta n = -\frac{\I e}{M\Omega^3}\frac{1}{1 + \I\nu/\Omega}\frac{\partial}{\partial z}\left(\frac{n_0(z)}{\varepsilon}\right)
\frac{\partial}{\partial y}H.
\end{equation}

\subsection{\label{Nonlin_curr}Nonlinear current.}
The nonlinear current is due to the first nonlinear correction
to the linear response velocities of electrons and
a scattering of an electromagnetic wave on the
distortion of the charge density profile produced by
another wave
\begin{equation}
\label{Current_NL_general}
\vec j_{NL} = e n_0(z) \vec v_{NL} + e\delta n \vec v_L.
\end{equation}
We introduce the nonlinear velocity $v_{NL}$ which can
be found from the following equation
$$
\frac{\partial \vec v_{NL}}{\partial t} = \vec v_L \times \left[ \nabla \times \vec v_L \right] +
\frac{\mu_0 e}{M}\vec v_L \times \vec H - \frac{1}{2}\nabla v_L^2 = - \frac{1}{2}\nabla v_L^2.
$$
Here we used a corollary of the Maxwell equations and
(\ref{Linear_v}) whence to within $O(\nu/\omega)$,
$$
\left[ \nabla \times \vec v_L \right] = -\frac{\mu_0 e}{M}\vec H.
$$
This means that only the dynamic pressure induced by the fields affects the plasma.

Finally, we have everything for the calculation of the first
term in right hand side of Ginzburg equation
(\ref{Ginzburg})
\label{curl_j_NL}
\begin{eqnarray}
\left(\nabla\times\vec j_{NL}\right)_x = \frac{\I e}{2\omega}\frac{d n_0(z)}{d z}\frac{\partial}{\partial y} v_L^2 +
e v_{z L}\frac{\partial}{\partial y}\delta n  -\\
- e v_{y L}\frac{\partial}{\partial z}\delta n - \frac{\mu_0 e^2}{M}\delta n \vec H.\nonumber
\end{eqnarray}
The detailed expression of equation's (\ref{Ginzburg})
right hand side can be found in Appendix~\ref{APP:Current_NL}.

\subsection{\label{Analytics_to}Analytic estimation. ``To the vehicle.''}
We would like to estimate the ratio
$$
\mu_S = \frac{S_S (z = R)}{S_0}
$$
of the fluxes of the squared scattered field to the squared incoming signal field and express it as a function of pump power flux $S_p$ measured in Watts per
square meter.

We can make an analytic estimation of the three-wave process
efficiency. The main contribution comes from the vicinity
of $z = 0$. The reason comes from the fact that the real part of dielectric susceptibility
(\ref{epsilon_z_Omega}) for the low frequency signal wave has a zero at this point.
It means that the nonlinear current on the right hand side of the Ginzburg equation
has a very sharp peak near $z=0$. A typical plot of right hand side is given in Figure \ref{RHS_to}.
This issue is discussed in more detail in Appendix \ref{APP:Current_NL_to}.

If we consider a high frequency pumping wave
we can use the plane wave approximation
$$
H_p(y,z,t) = H_p \E^{\I(-\omega_p t + k_p y - \kappa_p z)}.
$$
The low frequency signal wave can be written
$$
H_0(y,t) = H (y, z, t)|_{z=0} = H (z)|_{z=0} \E^{\I(-\omega t + k y)}.
$$
For the Stokes wave, whose frequency is higher than the
plasma frequency, one can use the following approximate
Ginzburg equation
\begin{equation}
\frac{\partial^2 H_S}{\partial z^2} + \kappa^2 H_S = f_S,
\end{equation}
where $f_S$ is calculated from the $\curl$ of the nonlinear current given in (\ref{curl_j_NL}).
To solve, we use the method of variation of constants.We find
\begin{eqnarray*}
H_S = C_1 \E^{\I\kappa_S z} + C_2 \E^{-\I\kappa_S z},\\
C_1' \E^{\I\kappa_S z} + C_2' \E^{-\I\kappa_S z} = 0,\\
C_1 (z) = \frac{1}{2\I \kappa_S}\int\limits_{-L}^{z} \E^{-i\kappa_S y} f_S (y) d y,\\
C_2 (z) = -\frac{1}{2\I \kappa_S}\int\limits_{z}^{R} \E^{i\kappa_S y} f_S (y) d y.\\
\end{eqnarray*}
One can say that $C_1$ is the amplitude of the Stokes wave propagating to the vehicle and $C_2$ is the amplitude of the anti-Stokes
wave propagating from the vehicle. The main contribution to $C_1 (R)$ arises from the vicinity of $z=0$,
where $f_S (z)$ is almost singular
\begin{eqnarray*}
C_1 (R) = \frac{1}{2\I \kappa_S}\int\limits_{-L}^{R} f_S
(y)\E^{-\I\kappa_S y} d y \simeq\\
\simeq \frac{1}{2\I\kappa_S}\int\limits_{-\infty}^{+\infty} f_S
(y)\E^{-\I\kappa_S y} d y.
\end{eqnarray*}
After some simple but tedious calculations (see Appendix \ref{APP:Current_NL_to}) one finds
\begin{equation}
C_1 (R) \simeq 2\pi\I\frac{e L}{Mc^2}\frac{1}{\varepsilon_0 c}\cos(2\theta)\sin(\phi)H_p H^*(0).\label{C_1_to_estimation}
\end{equation}
where $\theta$ is the pumping incident angle.

Details of these calculations are given in Appendix \ref{APP:Current_NL_to}. The angular dependence
of $H(0)$, which we call $\rho(\phi)$, can be calculated numerically by solving the homogeneous
Ginzburg equation.
In Fig.~\ref{RhoSinPhi}, we plot the product $\rho\sin\phi$ against $\phi$.
At the optimal value $\phi\simeq 0.5$, $\rho(\phi)\sin\phi \simeq 1/4$,
\begin{equation}
C_1 (R) \simeq \frac{\pi}{2}\I\frac{e L}{Mc^2}\frac{1}{\varepsilon_0 c}\cos 2\theta H_p H^*(-L).
\end{equation}
Using the expression
$S_p = |H_p|^2/(\varepsilon_0 c)$, one gets
\begin{eqnarray}
\label{estimation_to}
\mu_S = \left|\frac{C_1}{H}\right|^2
c\varepsilon_0\frac{S_p}{1 W m^{-2}}\simeq\nonumber\\
\simeq\frac{\pi^2}{4}\left(\frac{e L}{M c^2}\right)^2\frac{1}
{\varepsilon_0 c}\cos^2(2\theta)\frac{S_p}{1 W m^{-2}}.\label{Estimate_to}
\end{eqnarray}
For the optimal values of incidence angles ($\theta = 0$, $\phi \simeq 0.5$), the given plasma parameters and
$L \simeq 0.05 m$, one gets the following maximum value of the efficiency coefficient
\begin{equation}
\label{Estimate_to_number}
\mu_S \simeq 0.9\times10^{-11} \frac{S_p}{1 W m^{-2}}.
\end{equation}
This is consistent with what we obtain by direct numerical simulation.

\subsection{\label{Analytics_from}Analytic estimation. ``From the vehicle.''}
Equation (\ref{Ginzburg_equation}) can be rewritten in the following form
\begin{eqnarray}
\label{Ginzburg_rewritten_to}
\frac{\D}{\D z}\frac{1}{\varepsilon}\frac{\D H}{\D z} + \left(\frac{1}{\varepsilon_0}\frac{\omega^2}{c^2}
- \frac{k_0^2}{\varepsilon_0}\right)H =
\frac{\partial}{\partial z}\left(\frac{(\vec j_{NL})_y}{\varepsilon}\right) -\label{Ginzburg_from}\\
-\frac{1}{\varepsilon}\frac{\partial}{\partial y}(\vec j_{NL})_z.\nonumber
\end{eqnarray}
It is not too surprising that that the dominant contribution to the RHS of (\ref{Ginzburg_rewritten_to})
is the first term and arises from the neighborhood of $z=0$. Again, just as in the ``to the vehicle'' case,
the resonant layer acts as a transmitting antenna which will beam the message contained on the Stokes wave to
a distant receiver at frequency $\omega = \omega_p -\omega_S$. In Fig.~\ref{RHS_from} we verify that indeed
the dominant contribution comes from the first term on the RHS of (\ref{Ginzburg_rewritten_to}) and from the
neighborhood of $z=0$.
Hence we can get simple equation
for a very good approximation to the approximate particular solution of (\ref{Ginzburg_from}), namely
\begin{equation}
\frac{\D H}{\D z} = (\vec j_{NL})_y.
\end{equation}
The general solution is the following
\begin{equation}
H = C_1 \phi_1 (z) + C_2 \phi_2 (z) + \int\limits_{0}^{z} (\vec j_{NL})_y \D z,
\end{equation}
where $\phi_1 (z)$ and $\phi_2 (z)$ are solutions of the homogeneous part of equation (\ref{Ginzburg_from}), and
$\phi_1 (z)$ is bounded as $z \rightarrow R \gg 1$, $\phi_2 (z)$ is unbounded (exponentially)
at the vehicle. Thus $C_2 \simeq 0$. See Appendix~\ref{APP:Analytics_homogenious} for a discussion of solutions
to the homogeneous Ginzburg equation.

Using the boundary condition on the edge of the plasma ($z=-L$)
$$
\frac{\D H}{\D z}(-L) = -\I\kappa_0 H (-L),
$$
where $\kappa_0 = \frac{\omega_0}{c}\cos\phi$ is the $z$-component of wavevector of the outgoing low frequency signal
wave, and $\vec j_{NL} (-L) = 0$, one finds
\begin{equation}
C_1 = \frac{-\I\kappa_0}{\phi_1'(-L)+\I\kappa_0 \phi_1(-L)}\int\limits_{0}^{-L} (\vec j_{NL})_y \D z.
\end{equation}
Finally, for the magnetic field at $z=-L$ we find
\begin{equation}
H(-L) \simeq \frac{\phi_1'(-L)}{\phi_1'(-L)+\I\kappa_0 \phi_1(-L)}\int\limits_{0}^{-L} (\vec j_{NL})_y \D z.
\end{equation}
The function $(\vec j_{NL})_y$ oscillates with $z$ with wavenumber $\kappa_p - \kappa_S$. The lower
the wavenumber the more will be the contribution in the integral. This gives us a very simple optimal strategy
for the choice of pump and Stokes wave directions.
We should radiate both the Stokes and pumping waves in the desired direction
of the signal wave propagation. In this case we also have an exact compatibility with the boundary conditions
at $z=-L$.

If we consider the expression for $(\vec j_{NL})_y$ given in Appendix~\ref{APP:Current_NL}, we can see that in the
case $\omega_0 \ll \omega_S,\omega_p$ the first term (\ref{Current_NL_Y}) is the major one in the vicinity of
resonant layer. The resonant layer works like radiating antenna.

Using the simplified nonlinear current expression
and considering the pumping and Stokes waves as plane waves one finds
\begin{eqnarray}
H(-L) \simeq -\I\frac{e\omega_0^2 L \sin\phi}{2M\varepsilon_0 c^3 \omega_S \omega_p}\frac{1}{A}H_p H_S^*\times\\
\times\frac{\phi_1'(-L)}{\phi_1'(-L)+\I\kappa_0 \phi_1(-L)}\left(1-\frac{\E^{\I A\cos\phi} - 1}{A\cos\phi}\right).\nonumber
\end{eqnarray}
Where $A= L\omega_0/c$.

Using the solutions of the approximate homogeneous equations (\ref{GinzburgBessel}), we can estimate
$\phi_1'(z)/\phi_1(z)|_{z=-L} \simeq 1/L$. Thus for $\kappa_0 L = A\cos\phi \ll 1$, one finds
$$
H(-L) \simeq \frac{e\omega_0^2 L\sin\phi}{4M\varepsilon_0 c^3 \omega_S \omega_p}\frac{1}{A}H_p H_S^*.
$$
For the power density, we have
\begin{equation}
S = \frac{1}{32}\left(\frac{e L}{M c^2}\right)^2\frac{1}{\varepsilon_0 c}
\left(\frac{\omega_0^2}{\omega_S \omega_p}\right)^2\sin^2\phi S_S S_p.
\end{equation}
This result is quite clear from physical point of view. The larger $\phi$ is, the longer is the distance
over which the signal wave is generated in the plasma.

In our simulations, $A\simeq2.1$ and in this case we cannot use the simplified expression
given above. Instead we find,
\begin{eqnarray}
S = \frac{1}{8}\left(\frac{e L}{M c^2}\right)^2\frac{1}{\varepsilon_0 c}
\left(\frac{\omega_0^2}{\omega_S \omega_p}\frac{1}{A}\right)^2\times\nonumber\\
\times\tan^2\phi \left(1 - 2\frac{\sin(A\cos\phi)}{A\cos\phi}+2\frac{1-\cos(A\cos\phi)}{A^2 \cos^2\phi}\right)\times\label{estimation_from}\\
\times\frac{1}{1+C_{der}\cos^2\phi} S_S S_p.\nonumber
\end{eqnarray}
Here we introduced the coefficient $C_{der}=(\kappa_0 \phi_1/\phi_1')^2$ the value of which we obtain from
our numerics.

Finally, we find
\begin{eqnarray}
S_{12 GHz} = 1.2\times10^{-16}\tan^2\phi \left(1 - 2\frac{\sin(A\cos\phi)}{A\cos\phi}+\right.\nonumber\\
\left.+2\frac{1-\cos(A\cos\phi)}{A^2 \cos^2\phi}\right) \frac{1}{1+C_{der}\cos^2\phi} S_S S_p,\\
S_{18 GHz} = 2.0\times10^{-17}\tan^2\phi \left(1 - 2\frac{\sin(A\cos\phi)}{A\cos\phi}+\right.\nonumber\\
\left.+2\frac{1-\cos(A\cos\phi)}{A^2 \cos^2\phi}\right) \frac{1}{1+C_{der}\cos^2\phi} S_S S_p.
\end{eqnarray}
The subscripts refer to the frequencies of the onboard pump waves. Again, we find the magnitude and angular
dependence to be consistent with our numerical results.

\section{Numerical procedures and simulations.}
The equation we solve numerically in all cases is the Ginzburg equation
(\ref{Ginzburg}) including all terms on its right hand side.
The boundary conditions are given at $z = L_1 = -L-(L+R)$, in
the vacuum beyond the plasma edge and at $z=R$, the vehicle.

To solve this equation we use a ``sweep''-method described in detail in Appendix~\ref{APP:Numerical_method}.
The method was invented simultaneously in several places for work on classified topics in the middle of the last century.
In the Soviet Union, it was introduced by a group of L.\,D.~Landau (information from I.M. Khalatnikov)
(the first publication \cite{Sweep_Landau} appeared several years later due to obvious reasons)
and was developed to its modern form in \cite{Sweep_Gelfand}.

As the first step in the ``to the vehicle'' case we have to find the profile of the incoming magnetic field in the plasma.
We used an incident angle $\phi = 0.5$.
It will be shown later that this angle is an optimal value
but it is good for an initial evaluation of the possibility of communication. We consider the incoming signal as a monochromatic
plane wave of a given frequency $f_0 = 2 GHz$ and amplitude $H_0$. The current is equal to zero.
In this case, the boundary conditions are
\begin{eqnarray}
z &=& -L_1,\;\; \frac{\partial H}{\partial z} + \I \kappa_0 H = 2\I \kappa_0 H_0,\\
z &=& R,\;\; \frac{\partial H}{\partial z} = 0.
\end{eqnarray}
The resulting profile of the magnetic field is shown in Fig.~\ref{H_low_freq_to}. The profile of
$E_z (z)$ is shown in Fig.~\ref{EZ_low_freq_to}.
At the next stage, we consider an incident low frequency magnetic field profile as a source of distortion
of the plasma density profile
and take into account currents due to the presence of a pump wave. The pumping wave angle $\theta = 0.0$.
Our goal is to calculate the scattered field $H_S$ with frequency $\omega_S = \omega_p - \omega$.
In this case, the boundary conditions are
\begin{eqnarray}
z &=& -L_1,\; \frac{\partial H_S}{\partial z} + \I \kappa_{S} H = 0,\\
z &=& R,\; \frac{\partial H_S}{\partial z} = 0.
\end{eqnarray}	
The profiles of the magnetic fields $H_S$
for two different pumping frequencies are shown in Figures~\ref{H_high_freq_to_12GHz} and \ref{H_high_freq_to_18GHz}. We note that the resonant layer $z=0$ acts as if it were a source.

In the ``from the vehicle'' case we calculate the magnetic field of the low frequency wave generated by plane
pump and Stokes waves.
Following the optimal strategy in this case, described in the analytic part of the paper, we take
all angles equal to each other $\phi=\theta=\pi/4$.
In this case, the boundary conditions are
\begin{eqnarray}
z &=& -L_1,\; \frac{\partial H}{\partial z} + \I \kappa_{0} H = 0,\\
z &=& R,\; H = 0.
\end{eqnarray}
Here $H(z)$ is the magnetic field of the signal wave with frequency $\omega = \omega_p - \omega_S$.
The boundary condition at $z = R,\; H = 0$, gives us the worst of all cases by definition.

The low frequency magnetic fields
for two different pumping frequencies are shown in Figs.~\ref{H_low_freq_from_12GHz} and \ref{H_low_freq_from_18GHz}.

We tested the robustness of the code by allowing for both finite and zero conductivity of the vehicle surface
in the ``to the vehicle case''. During the simulation in the ``from the vehicle'' case we also redid the
simulation with the
derivative of the magnetic field at the vehicle equal to zero. In all the cases, the influences of the differing boundary
conditions were negligible.

In the ``to the vehicle'' case, it is convenient to introduce the function $\mu_S$ as the ratio
$$
\mu_S = \frac{S_S (z = R)}{S_0}
$$
of the scattered field flux to the incoming signal flux and express it as a function of pump flux $S_p$ measured in Watts per
square meter.
We found
$$
\omega_p = 2\pi * 12GHz, \; \max(\mu_S)  \simeq 2.2\times10^{-12} \frac{S_p}{1Wm^{-2}},
$$
$$
\omega_p = 2\pi * 18GHz, \; \max(\mu_S)  \simeq 0.63\times10^{-11} \frac{S_p}{1Wm^{-2}}.
$$
These results are in a good agreement with the analytic estimation (\ref{Estimate_to_number}).
Any difference is due to the fact that the pumping frequency is not sufficiently high to neglect
the plasma frequency. The reason we used these frequencies and not much higher ones was that they are available
on standard microwave equipment and devices.

In the ``from the vehicle'' case, we calculate the ratio
$$
\mu = \frac{S_{out} (z = -L)}{S_S S_p}
$$
of the output signal flux to the product of the pump and Stokes fluxes and express it as a function of the optimal angle.

We found
$$
\omega_P = 2\pi * 12GHz, \; \max(\mu) \simeq 1.8\times10^{-16} \frac{1}{1Wm^{-2}},
$$
$$
\omega_P = 2\pi * 18GHz, \; \max(\mu) \simeq 3.0\times10^{-17} \frac{1}{1Wm^{-2}}.
$$

In order to investigate the dependence of the result on the angles $\phi, \theta_p, \theta_S$, we calculated $\mu$
for various different choices. The results are shown in Figs.~\ref{Theta_Phi_to_12GHz}-\ref{Theta_Phi_from_18GHz}.

As one can see, in the ``to the vehicle'' case we have a very good agreement between the analytically estimated
angular dependence (\ref{estimation_to}) and the numerical results. Namely, we have a maximum at pumping angles
close to $\theta = 0$ and the efficiency coefficient $\mu_S$ goes to zero at the vicinity of $\theta = \pi/4$ in
a agreement with the $\cos(2\theta)$ dependence. So we can formulate a simple rule: in order to get the
best possible performance, send the pump wave in a direction perpendicular to the plasma edge surface.

In the ``from the vehicle'' case, the situation is even simpler. As it was shown in Section \ref{Analytics_from}
the power conversion is optimal if we radiate both the pump and Stokes waves in the direction of
the desired signal wave propagation. The estimated angular dependence (\ref{estimation_from}) can be
fitted with good accuracy to the numerical results using only one tuning coefficient $C_{der}$. It is shown
that this coefficient weakly depends on the pumping frequency.

\section{Conclusion and discussion.}
Let us now discuss the practical usage of this approach for receiving at and transmitting from the vehicle.
For the ``to the vehicle'' case we consider the problem of receiving even GPS signals. Let us estimate
the resulting attenuation coefficient.
Given a pump waveguide aperture of $3cm\times 3cm$ and a pump power of $3kW$, this gives $S_p =
3.3\times10^{6}Wm^{-2}$. One can use the pulse regime. In this case, even for pulses $10^{-3}s$ long,
every pulse still contains more than $10^6$ periods of the low frequency signal and we can get
much higher power flux $S_p^{pulse} = 3.3\times 10^{9}Wm^{-2}$.
It gives us the attenuation coefficients $\mu_S S_{p}^{pulse}$
\begin{eqnarray*}
\mu_S S_p^{pulse} \simeq 0.73\times 10^{-2},\;\omega_p &=& 2\pi * 12GHz.\\
\mu_S S_p^{pulse} \simeq 2.1\times 10^{-2},\;\omega_p &=& 2\pi * 18GHz.
\end{eqnarray*}
The usual level of a GPS signal at the Earth surface is about $-127.5 dBm$ (1 Decibel per milliwatt is equal
to $1dBm = 10 \log_{10}(P/1mW)$). Indoors, one must use high sensitivity
GPS receivers. Many general purpose chipsets have been available for several years. Presently,
the market offers sensitivities $-157.5 dBm$ (for example \cite{Fujitsu}).
Using the definition of $dBm$ one can see, that it is possible to receive a signal with an attenuation about
$10^{-3}$. Also it is possible to use a much bigger
antenna on the vehicle than in the case of a handheld device. In this case, it is even possible to receive a signal
using the continuous rather than pulsed regime for a klystron pump. So even at the angles far from
optimal,  one can receive GPS signals. Further, we used maximum value of the plasma thickness.
If the plasma sheath is thinner, the angular dependence is broader.

Some characteristics of klystron amplifiers available on the open market are given in Table~\ref{Table_1} \cite{NEC}.

In the ``from the vehicle'' case, because of sensitive land based receivers, all we need is to have a reasonable signal.
Let us estimate an incoming power on some land based antenna.
First of all, for any real antenna we have to take into account the decrease of a signal due to diffraction broadening.
If the diameter of the land-based antenna (Figure \ref{antenna}) is $D_0$,
the diameter of the signal flux after some long distance $l$ will be
\begin{equation}
D(l) \simeq \frac{l\lambda}{2 D_0}.
\end{equation}
It means that if we have power flux at an antenna $S_A$, the power flux at the edge of the plasma after a
distance $l$ will be
\begin{equation}
S_0 \simeq S_A \left(\frac{2D_0^2}{l\lambda}\right)^2.
\end{equation}
For example, for an antenna of the diameter equal to 5m, after 100km
$$
S_0 \simeq 1.1\times10^{-5} S_A.
$$
Now one can calculate the sensitivity of the receiver needed.
Let us suppose that the signal beam outgoing from the vehicle
has diameter $D_0 = 1m$, signal frequency $f = 2 GHz$ and corresponding wave length
$\lambda = 1.5\times 10^{-1}m$, the land based antenna has a diameter $D_{LB} = 5m$ and is situated at a distance $l = 100 km$.
Using the previous results for diffraction, the pumping klystrons' powers from the table above and the expression
$S_{out}=\mu S_pS_S$, one can get for the power on the land based receiver
\begin{equation}
S_{LB} \simeq S_{out} \left(\frac{2D_0^2}{l\lambda}\right)^2 = 1.8\times 10^{-8} S_{out}.
\end{equation}
We now list for two different frequencies, the corresponding powers in Watts at the receiving antenna.
\begin{eqnarray*}
\omega_P &=& 2\pi * 12GHz,\\
P_A &\simeq& 1.8\times10^{-8}*
1.8\times10^{-16}*9\times10^{6}Wm^{-2}* 25m^2\\
&\simeq& 0.73\times 10^{-15} W;\\
\omega_P &=& 2\pi * 18GHz,\\
P_A &\simeq& 1.8\times10^{-8}*
3.0\times10^{-17}*4\times10^{12}Wm^{-2}* 25m^2\\
&\simeq& 0.54\times 10^{-17} W.
\end{eqnarray*}

The GPS receiver mentioned above has a sensitivity about $-160dBm \simeq 10^{-19}W$.
Even with such a modest size of the
antenna and ordinary klystrons one can receive the signal at almost any angle.

As a final remark one can conclude that proposed method for communication with and from the supersonic vehicle is
realistic even using standard devices available on the open market.

In a future work, there are several additional points we would like to consider further. First,
it might be worthwhile to discuss the effect of the plasma density profile at $z<0$ on the wave
interaction. One may expect that the transition will be much narrower than $5\,cm$, because of the shock formation.
But shock waves take place not in the plasma but in air. The air and plasma densities
are not connected directly. The plasma density is defined by the level of ionization
which is governed by a density distribution in accordance with the Saha
equation. In a real plasma sheath the characteristic temperature is much
less then the ionization potential, and a typical level of ionization is
low ($10^{-6}$-$10^{-5}$). Under this conditions the plasma density depends on the
temperature dramatically. The temperature jumps inside the shock then
grows gradually toward the vehicle. Still, just after the shock it is too
low to provide a strong ionization, and the most essential increase in
ionization takes place far behind the shock. For this reason we can neglect
the jump of plasma density inside the shock wave and treat it as smooth.
In this case an approximation by a linear function seems to be reasonable.
Let us note that there is no blackout if the plasma sheath is as thick as
$5\,cm$ and the plasma density is as low as $10^{18}\,m^{-3}$. In this case, the incident
wave can reach the vehicle due to the skin-effect. Anyway, our numerical code
can be used for an arbitrary density profile.

Another question is the following: will the shock and the flow behind it suffer from hydrodynamic instabilities?
Actually, hydrodynamic-type instabilities as well as hydrodynamic turbulence are
slow processes in our time-scale. We can treat the plasma sheath as
``frozen''. The distortion of the density profile, although frozen, can slightly change the results.

It could look quite surprising that collision frequency $\nu$ drops out from the final results after integration
over the spatial variable. This is a common mathematical trick, working perfectly
as long as ${\nu}/{\omega}\rightarrow 0$. In our case we have
${\nu}/{\omega}\simeq 10^{-1} - 10^{-2}$. In fact, the shape of the resonant layer can be distorted by nonlinear
effects. They are essential if the ongoing signal is powerful enough.
At very small $\nu$, the resonant peak would become very narrow.
Could then the electron thermal motion start affecting the structure of the resonance?
This question is very important. A resonant layer cannot be
thinner than the Debye length. In our case $r_d\simeq 10^{-4}\,cm$
while the thickness of the resonant layer is $l\simeq L\cdot({\nu}/{\omega})^2\simeq 10^{-1}-10^{-2}cm$.
Thus the influence of the temperature can be neglected.
In the case of much thinner resonant layer this phenomenon should be taken into
account and considered separately.
Anyway, such influence could lead only to a radiation of Langmuir waves from the resonant area.
But this effect is nothing but additional dissipation, dropping out from the
final formulae.

When the signals come from the external source, what would be the role of the $s$-polarized component? One can
expect that, for some orientations of the vehicle, the s-component will be dominant, and the resonance will
disappear.This could somewhat reduce the efficiency.
In our idealized model (the vehicle is an infinite wall) there is
a difference between $s$ and $u$ polarizations. For a real vehicle
such a difference could be important. This is a subject for future study.

As one can see there are still a lot of open questions. The physics of the problem is very rich.
Our present work may be best described as a "proof of concept" research from which we can make some orders of
magnitude estimates and gain some insights. The results can be made more accurate by including some of the
physical effects discussed above.

\section{Acknowledgments} 
This work was  supported by AFOSR contract number~FH\,95500410090.

A.O. Korotkevich was supported by RFBR grant 06-01-00665-a, the Programme 
``Nonlinear dynamics and solitons'' from the RAS Presidium and ``Leading 
Scientific Schools of Russia'' grant.

\appendix

\section{\label{APP:Analytics_homogenious}Analytic solutions of the Ginzburg equation in some special cases.}
By neglecting $\vec j_{NL}$, we obtain the linear Ginzburg
equation. It takes an especially simple form if $\Omega = \omega$,
$\nu/\omega~=~0$ and $H\sim \E^{\I k y}$. In this case
$\varepsilon/\varepsilon_0 = - z/L$ and equation
(\ref{Ginzburg}) is
\begin{equation}
\label{GinzburgLinear}
\frac{\D^2 H}{\D z^2} - \frac{1}{z}\frac{\D H}{\D z} -
\left(\frac{z}{\Lambda^3} + k^2\right)H = 0.
\end{equation}
Here
\begin{equation}
\label{Lambda}
\Lambda = \left(\frac{c^2}{\omega_L^2}(L+R)\right)^{1/3} =
\left(\frac{c^2}{\omega^2} L\right)^{1/3},
\end{equation}
and $\Lambda$ is another length. In our case
$\omega_L \simeq 2\pi\times 9 GHz$, $R+L = 1m$ and
$\Lambda=0.03m \simeq L$.

One can introduce the dimensionless variable $\xi = z/\Lambda$.
Then equation (\ref{GinzburgLinear}) simplifies to,
\begin{equation}
\label{GinzburgLinearDimensionless}
\frac{\D^2 H}{\D \xi^2} - \frac{1}{\xi}\frac{\D H}{\D \xi} -
\left(\xi + \alpha^2\right)H = 0.
\end{equation}
Here $\alpha^2 = \Lambda^2 k^2$ is a dimensionless constant.

Equation (\ref{GinzburgLinear}) has two linearly independent
solutions $\phi_1$, $\phi_2$. We assume
$$
\phi_1 \rightarrow 0,\; \phi_2 \rightarrow \infty,\;
\mbox{at}\; z \rightarrow \infty. 
$$
The Wronskian of these solutions is proportional to
$\varepsilon/\varepsilon_0$. We can put
\begin{equation}
\label{Wronskian}
W = \{\phi_1, \phi_2\} = \phi_1'\phi_2 - \phi_2'\phi_1 =
-\frac{z}{L}.
\end{equation}
It means that
$$
W|_{z=-L} = 1.
$$
Equation (\ref{GinzburgLinearDimensionless}) cannot be solved
in terms of any known special functions. In the ``outer'' area
$|\xi| \gg \alpha^2$ it reduces to the form
\begin{equation}
\label{GinzburgAiry}
\frac{\D^2 H}{\D \xi^2} - \frac{1}{\xi}\frac{\D H}{\D \xi} -
\xi H = 0.
\end{equation}
One can check that equation (\ref{GinzburgAiry}) can be solved
in terms of the Airy functions $\Ai$ and $\Bi$. Namely,
\begin{eqnarray}
\phi_1 = a_1 \Ai'(\xi) \sim \frac{a_1}{2\sqrt{\pi}}\xi^{1/4}
\E^{-2/3 \xi^{3/2}},\\
\phi_2 = b_1 \Bi'(\xi) \sim \frac{b_1}{\sqrt{\pi}}\xi^{1/4}
\E^{2/3 \xi^{3/2}},\;\mbox{at}\;\xi\rightarrow\infty.\nonumber
\end{eqnarray}
From (\ref{Wronskian}) one gets
\begin{equation}
a_1 b_1 = \frac{\pi \Lambda^2}{L}.
\end{equation}

In the ``inner'' area $|\xi| \ll \alpha^2$, the equation
(\ref{GinzburgLinearDimensionless}) is reduced to the form
\begin{equation}
\label{GinzburgBessel}
\frac{\D^2 H}{\D \xi^2} - \frac{1}{\xi}\frac{\D H}{\D \xi} -
\alpha^2 H = 0.
\end{equation}
Equation (\ref{GinzburgBessel}) can be solved in terms of
Bessel functions \cite{Ginzburg_bib}. Two linearly independant
solutions of equation (\ref{GinzburgBessel}) $\psi_1$,
$\psi_2$ behave in neighborhood of $\xi = 0$ as follows
\begin{eqnarray}
\psi_1&=&1 + \frac{\alpha^2}{2}\xi^2 \left(\log \xi - \frac{1}{2}\right) + ...\\
\psi_2&=&\xi^2 + \frac{\alpha^2}{8}\xi^4 + ...\nonumber
\end{eqnarray}
Both solutions, which are some linear combinations of $\psi_1$,
$\psi_2$ are bounded. Thus the magnetic field has no singularity at
$z=0$.

\section{\label{APP:Current_NL}Right hand side of the Ginzburg equation.}
\subsection{General case.}
Consider the Ginzburg equation for a wave
$$
H_3 (y,z,t) = H_3 (z) \E^{-\I\omega_3 t + \I k_3 y},
$$
and calculate the right hand side of (\ref{Ginzburg}) in terms if the fields $H_1(y,z,t), H_2(y,z,t)$.
In the ``to the vehicle'' case, $H_1$ will represent the pump wave, $H_2$ the signal wave and $H_3$
the Stokes wave.
In the ``from the vehicle'' case, $H_3$ will be the signal and $H_1$ and $H_2$ the pump and signal carrying
Stokes waves respectively. In all cases $\omega_3 = \omega_2 - \omega_1,\; k_3 = k_2 - k_1$. We find
\begin{widetext}
\begin{eqnarray}
\label{Current_NL_vortex_x}
\left[\nabla\times\vec j_{NL}(H_1,H_2,k_1,k_2,k_3,\omega_1,\omega_2,\omega_3)\right]_x =\nonumber\\
= -\frac{e^3 n_0'(z) k_3}{2M^2 \omega_3 \left(1+\I\nu/\omega_3\right)}
\left( \frac{1}{\varepsilon_1^* \left(1 - \I\nu/\omega_1\right)\omega_1^2 \varepsilon_2 \left(1 + \I\nu/\omega_2\right)\omega_2^2}
\frac{\partial H_1^*}{\partial z}\frac{\partial H_2}{\partial z} +\right.\\
\left.+ \frac{k_1 k_2}{\varepsilon_1^* \left(1 - \I\nu/\omega_1\right)\omega_1^2
\varepsilon_2 \left(1 + \I\nu/\omega_2\right)\omega_2^2}H_1^* H_2\right) +\nonumber\\
+ \frac{e^3}{M^2}\left(\frac{k_2k_1^2}
{\left(1 - \I\nu/\omega_1\right)\omega_1^3 \varepsilon_2 \left(1 + \I\nu/\omega_2\right)\omega_2^2}\frac{\partial}{\partial z}\left(\frac{n_0(z)}{\varepsilon_1^*}\right)\right.+\nonumber\\
+\left.\frac{k_2^2k_1}
{\left(1 - \I\nu/\omega_1\right)\omega_2^3 \varepsilon_1^* \left(1 + \I\nu/\omega_2\right)\omega_1^2}\frac{\partial}{\partial z}\left(\frac{n_0(z)}{\varepsilon_2}\right)\right)H_1^*H_2\\
+ \frac{e^3}{M^2}\left(\frac{1}{\omega_2^2 \varepsilon_2\left(1+\I\nu/\omega_2\right)}\frac{\partial H_2}{\partial z}
\frac{k_1}{\omega_1^3\left(1-\I\nu/\omega_1\right)}\frac{\partial }{\partial z}\left(H_1^*\frac{\partial }{\partial z}\left(\frac{n_0(z)}{\varepsilon_1^*}\right)\right)+\right.\nonumber\\
+\left.\frac{1}{\omega_1^2\varepsilon_1^*\left(1-\I\nu/\omega_1\right)}\frac{\partial H_1^*}{\partial z}
\frac{k_2}{\omega_2^3\left(1+\I\nu/\omega_2\right)}\frac{\partial }{\partial z}\left(H_2\frac{\partial }{\partial z}\left(\frac{n_0(z)}{\varepsilon_2}\right)\right)\right)-\\
-\frac{\mu_0 e^3}{M^2}\left(\frac{k_2}{\omega_2^3\left(1+\I\nu/\omega_2 \right)}\frac{\partial}{\partial z}\left(\frac{n_0(z)}{\varepsilon_2}\right) +
\frac{k_1}{\omega_1^3\left(1-\I\nu/\omega_1\right)}\frac{\partial}{\partial z}\left(\frac{n_0(z)}{\varepsilon_1^*}\right)\right)H_1^* H_2.
\end{eqnarray}
Using formulae (\ref{Current_NL_general}),
\begin{eqnarray}
\left(\vec j_{NL}(H_1,H_2,k_1,k_2,k_3,\omega_1,\omega_2,\omega_3)\right)_y =\nonumber\\
= \frac{e^3 n_0(z) k_3}{2M^2 \omega_3 \left(1+\I\nu/\omega_3\right)}
\left( \frac{1}{\varepsilon_1^* \left(1 - \I\nu/\omega_1\right)\omega_1^2 \varepsilon_2 \left(1 + \I\nu/\omega_2\right)\omega_2^2}
\frac{\partial H_1^*}{\partial z}\frac{\partial H_2}{\partial z} +\right.\label{Current_NL_Y}\\
\left.+ \frac{k_1 k_2}{\varepsilon_1^* \left(1 - \I\nu/\omega_1\right)\omega_1^2
\varepsilon_2 \left(1 + \I\nu/\omega_2\right)\omega_2^2}H_1^* H_2\right) -\nonumber\\
- \frac{e^3}{M^2}\left(\frac{1}{\omega_2^2 \varepsilon_2\left(1+\I\nu/\omega_2\right)}\frac{\partial H_2}{\partial z}
\frac{k_1}{\omega_1^3\left(1-\I\nu/\omega_1\right)}H_1^*\frac{\partial }{\partial z}\left(\frac{n_0(z)}{\varepsilon_1^*}\right)+\right.\nonumber\\
+\left.\frac{1}{\omega_1^2\varepsilon_1^*\left(1-\I\nu/\omega_1\right)}\frac{\partial H_1^*}{\partial z}
\frac{k_2}{\omega_2^3\left(1+\I\nu/\omega_2\right)}H_2\frac{\partial }{\partial z}\left(\frac{n_0(z)}{\varepsilon_2}\right)\right).
\end{eqnarray}
\subsection{\label{APP:Current_NL_to}Approximate right hand side. ``To the vehicle'' case.}
In the ``to the vehicle'' case, the main contribution comes from the terms containing poles
\begin{eqnarray}
\left[\nabla\times\vec j_{NL}(H,H_p,k,k_p,k_S,\omega,\omega_p,\omega_S)\right]_x \simeq\\
\simeq \frac{e^3}{M^2}\frac{k_p k^2}{\left(1 - \I\nu/\omega_p\right)\omega^3 \varepsilon_p \left(1 + \I\nu/\omega_p\right)\omega_p^2}\frac{\partial}{\partial z}\left(\frac{n_0(z)}{\varepsilon^*}\right)H^* H_p+\nonumber\\
+ \frac{e^3}{M^2}\frac{1}{\omega_p^2 \varepsilon_p\left(1+\I\nu/\omega_p\right)}\frac{\partial H_p}{\partial z}
\frac{k}{\omega^3\left(1-\I\nu/\omega\right)}\frac{\partial }{\partial z}\left(H^*\frac{\partial }{\partial z}\left(\frac{n_0(z)}{\varepsilon^*}\right)\right)-\nonumber\\
-\frac{\mu_0 e^3}{M^2}\frac{k}{\omega^3\left(1-\I\nu/\omega\right)}\frac{\partial}{\partial z}\left(\frac{n_0(z)}{\varepsilon^*}\right)H^* H_p.\nonumber\\
\end{eqnarray}
\end{widetext}
Assume that the high frequency pumping wave remains undisturbed. Then
$$
H_p (y,z,t) = H_p \E^{\I(k_p y - \kappa_p z -\omega t)},
$$
we find
$$
f_S (z) = - \left[\nabla\times\vec j_{NL}\right]_x \E^{-\I\kappa_p z}
$$
and
\begin{eqnarray*}
C_1 (R) = \frac{1}{2\I \kappa_S}\int\limits_{-L}^{R} f_S
(y)\E^{-\I\kappa_S y} d y \simeq\\
\frac{1}{2\I \kappa_S}\int\limits_{-\infty}^{+\infty} f_S
(y)\E^{-\I\kappa_S y} d y.
\end{eqnarray*}
After several integrations by parts in the second term of $\left[\nabla\times\vec j_{NL}\right]_x$,
taking into account $k_S = k_p - k$, one finds
\begin{eqnarray}
C_1 (R) \simeq \frac{-ik}{2\kappa_S}\frac{e^3 H_p H^*}{M^2\varepsilon_0\omega_p^2 \omega^3}
\left(k_p (k_p - k_S) + (\kappa_p +\right.\\
\left.\kappa_S)\kappa_p -
\frac{\omega_p^2}{c^2}\right)\int\limits_{-\infty}^{+\infty}
\frac{\partial}{\partial z}\left(\frac{n_0(z)}{\varepsilon^*}\right)\E^{-\I(\kappa_S + \kappa_p)z}\D z.\nonumber
\end{eqnarray}
For most pumping angles, and using the fact that $\omega_p \gg \omega$, one can substitute $\omega_S \simeq \omega_p$
and consider incidence angles of pumping and Stokes wave to be close in absolute value. Following
Fig.~\ref{plasma_response_fig} the pumping incidence angle is $\theta$ and low frequency signal
incidence angle is $\phi$.
\begin{eqnarray}
C_1 (R) \simeq \frac{-ik}{2\kappa_S}\frac{e^3 H_p}{M^2\varepsilon_0\omega_p^2 \omega^3}
\frac{\omega_p^2}{c^2}\cos(2\theta)\\
\int\limits_{-\infty}^{+\infty}
\frac{\partial}{\partial z}\left(\frac{n_0(z)}{\varepsilon^*}\right)H^*\E^{-\I(\kappa_S + \kappa_p)z}\D z.\nonumber
\end{eqnarray}
Using integration by parts once more one can get
\begin{eqnarray}
C_1 (R) \simeq \frac{(\kappa_p + \kappa_S)}{2\kappa_S}\frac{e^3 H_p}{M^2\varepsilon_0 c^3 \omega^2}
\cos(2\theta)\sin(\phi)\times\\
\times\int\limits_{-\infty}^{+\infty}
\left(\frac{n_0(z)}{\varepsilon^*}\right)H^*\E^{-\I(\kappa_S + \kappa_p)z}\D z.\nonumber
\end{eqnarray}
Calculating this integral by residues and taking into account
\begin{eqnarray*}
n_0 (z)\simeq n_0 (z=0) = n_0 \frac{L}{L+R} = n_0\frac{\omega^2}{\omega_L^2},\\
\frac{\varepsilon_0}{\varepsilon} = -\frac{L}{z+i\delta},\;
\delta = \I\frac{\nu}{\omega}L,
\end{eqnarray*}
we finally get
\begin{eqnarray*}
C_1 (R) \simeq 2\pi\I\frac{e^3 n_0 L}{M^2\varepsilon_0^2 c^3\omega_L^2}\cos(2\theta)\sin(\phi)H_p H^*(0)=\nonumber\\
=2\pi\I\frac{e L}{Mc^2}\frac{1}{\varepsilon_0 c}\cos(2\theta)\sin(\phi)H_p H^*(0).
\end{eqnarray*}

\section{\label{APP:Numerical_method}Numerical method.}
Here we briefly present formulae for the ``sweep'' method in a very general way following the approach
given in \cite{Fedorenko}.
\subsection{Reformulation of a problem on a grid.}
Consider the ODE
\begin{equation}
\label{num_equation}
p(x) \frac{d^2 y}{dx^2} + q(x)\frac{dy}{dx} + r(x) y = f(x),
\end{equation}
in the region $0 < x < L$ with boundary conditions
\begin{eqnarray}
\alpha \frac{dy}{dx} + \beta y|_{x=0} = \gamma,\label{num_boundary}\\
\alpha_1 \frac{dy}{dx} + \beta_1 y|_{x=L} = \gamma_1\nonumber.
\end{eqnarray}
We use for (\ref{num_equation}) a second order finite difference scheme
on a grid of $(N+1)$-nodes ($y_0 = y(0), y_N = y(L)$) with constant step $h$
\begin{equation}
\label{num_equation_diff}
p_n \frac{y_{n+1} - 2 y_n + y_{n-1}}{h^2} + 
q_n \frac{y_{n+1} - y_{n-1}}{2h} + r_n y_n = f_n.
\end{equation}
This equation is only valid for inner nodes of the grid.

The boundary conditions take the form
\begin{eqnarray}
\alpha \frac{y_1 - y_0}{h} + \beta y_0 = \gamma,\label{num_boundary_diff}\\
\alpha_1 \frac{y_{N} - y_{N-1}}{h} + \beta_1 y_N = \gamma_1\nonumber.
\end{eqnarray}
We can rewrite equation (\ref{num_equation_diff}) as
\begin{eqnarray}
a_n y_{n-1} - b_n y_n + c_n y_{n+1} = d_n,\label{num_matrix_inner}\\
c_n = \frac{p_n}{h^2} + \frac{q_n}{2h},\;
a_n = \frac{p_n}{h^2} - \frac{q_n}{2h},\nonumber\\
b_n = \frac{2p_n}{h^2} - r_n = a_n + c_n - r_n,\;
d_n = f_n.\nonumber
\end{eqnarray}
In the same way for (\ref{num_boundary_diff}),
one finds
\begin{eqnarray}
- b_0 y_0 + c_0 y_1 = d_0,\label{num_matrix_left}\\
b_0 = \frac{\alpha}{h} - \beta, c_0 = \frac{\alpha}{h}, d_0 = \gamma\nonumber\\
a_N y_{N-1} - b_N y_N = d_N,\label{num_matrix_right}\\
a_N = -\frac{\alpha_1}{h}, b_N = -\frac{\alpha_1}{h} - \beta_1, d_N = \gamma_1\nonumber\\
\end{eqnarray}
The result is a tridiagonal matrix $\left((N+1)\times(N+1)\right)$ equation for $a,b,c$.

\subsection{``Sweep''-method.}
The solution of the linear system of equations with tridiagonal matrix is well described in numerous sources (for instance
\cite{Numerical_Recipes}).
It can be shown that one can find a solution in the following form
\begin{equation}
\label{sweep_relation}
y_{n-1} = P_n y_n + Q_n.
\end{equation}
From the left boundary, we have from (\ref{num_matrix_left}), that
$$
y_0 = \frac{c_0}{b_0} y_1 + \frac{d_0}{b_0}.
$$
In this case
\begin{eqnarray}
\label{initial_for_PQ}
P_1 = \frac{c_0}{b_0},\; Q_1 = -\frac{d_0}{b_0}.
\end{eqnarray}
Next we derive a recurrence relation for $P_n$ and $Q_n$. After substituting
(\ref{sweep_relation}) in (\ref{num_matrix_inner}), we find
$$
y_n = \frac{c_n}{b_n - a_n P_n} y_{n+1} + \frac{a_n Q_n - d_n}{b_n - a_n P_n}.
$$
Then, comparing with (\ref{sweep_relation}), we see that
\begin{equation}
\label{direct_sweep_recurring}
P_{n+1} = \frac{c_n}{b_n - a_n P_n},\; Q_{n+1} = \frac{a_n Q_n - d_n}{b_n - a_n P_n}.
\end{equation}
Using the initial values (\ref{initial_for_PQ}) and the recurring relations
(\ref{direct_sweep_recurring}), one can get all
$P_n, Q_n$ coefficients up to $n=N$ (``direct sweep'' from left to right).

Than we use second (right) boundary condition (N-th equation)
$$
a_N y_{N-1} - b_N y_N = d_N\;\;
\mbox{and}\;\; y_{N-1} = P_N y_N + Q_N.
$$
Immediately one finds
\begin{equation}
\label{y_initial}
y_N = \frac{d_N - a_N Q_N}{a_N P_N - b_N}
\end{equation}
Finally, performing a recurrent ``backward sweep'' (from right to left), using the already known $P_n, Q_n$,
``sweep''-relations (\ref{sweep_relation}) and the initial condition (\ref{y_initial}),
we get values for all $y_n$.

\newpage

\begin{table}[ht!]
\begin{tabular}{|c|c|c|c|}
\hline
Model & Frequency (GHz) & Power (kWatt) & Mass (kg) \\
\hline
LD4595 & 14.0-14.5 & 3 & 40 \\
\hline
LD7126 & 17.3-18.4 & 2 & 27 \\
\hline
\end{tabular}
\caption{\label{Table_1}Characteristics of some klystrons available on the open market.}
\end{table}

\newpage

\begin{figure}[ht!] 
\caption{\label{plasma_response_fig}$\omega_L (z_r) = \omega\cos\phi, \omega_L (0) = \omega.$ If the thickness of the plasma sheath is
equal to $L+R = 1m$, the signal frequency $f = 2 GHz$ and the plasma frequency $f_L \simeq 9 GHz$ then $L \simeq 5cm$ and $R \simeq 95cm$.} 
\end{figure}

\begin{figure}[ht!] 
\caption{\label{from_concept} The concept for communication from the vehicle. Although drawn in such a way
that the angles of pumping, Stokes and signal waves are different, the optimal configuration is when
all angles are the same, i.e. Stokes and pump waves are generated in the same direction as the target of the
desired low frequency signal.} 
\end{figure}

\begin{figure}[ht!] 
\caption{\label{RHS_to}The typical right hand side (absolute value)
of the Ginzburg equation in the ``to the vehicle'' case.
Logarithmic scale. One can see that the main contribution comes from the region of the point $z = 0$.} 
\end{figure}

\begin{figure}[ht!]
\caption{\label{RhoSinPhi}Dependence of $C_1(R)$ on the signal incidence angle $\phi$.}
\end{figure}

\begin{figure}[ht!] 
\caption{\label{RHS_from}The full right hand side (absolute value)
of the Ginzburg equation in the ``from the vehicle'' case (solid line) together with the
expression used in the approximation (dashed line).
Logarithmic scale. Before the vicinity of point $z = 0$ almost no forcing is present. Almost all contributions
in the vicinity of the resonant point comes from the term used in the approximation. In the propagation region ($z < 0$), the approximation slightly underestimates the right hand side.}
\end{figure}

\begin{figure}[ht!] 
\caption{\label{H_low_freq_to} Incoming signal magnetic field profile.} 
\end{figure}

\begin{figure}[ht!] 
\caption{\label{EZ_low_freq_to} Incoming signal electric field (z-component) profile.} 
\end{figure}

\begin{figure}[ht!] 
\caption{\label{H_high_freq_to_12GHz} Magnetic field profile of the Stokes wave. Pumping frequency 12GHz.}  
\end{figure}

\begin{figure}[ht!]
\caption{\label{H_high_freq_to_18GHz} Magnetic field profile of the Stokes wave. Pumping frequency 18GHz.} 
\end{figure}

\begin{figure}[ht!] 
\caption{\label{H_low_freq_from_12GHz} Generated low frequency magnetic field. Pumping frequency 12GHz.}  
\end{figure}

\begin{figure}[ht!]
\caption{\label{H_low_freq_from_18GHz} Generated low frequency magnetic field. Pumping frequency 18GHz.} 
\end{figure}

\begin{figure}[ht!]
\caption{\label{Theta_Phi_to_12GHz} Dependence of power conversion efficiency coefficient $\mu_S$ on angles. ``To the vehicle''. Pumping frequency 12GHz.}
\end{figure}

\begin{figure}[ht!]
\caption{\label{Theta_Phi_to_18GHz} Dependence of power conversion efficiency coefficient $\mu_S$ on angles. ``To the vehicle''. Pumping frequency 18GHz.}  
\end{figure}

\begin{figure}[ht!] 
\caption{\label{Theta_Phi_to_12GHz_small} Dependence of the power conversion efficiency coefficient $\mu_S$ on several pumping angles, in the ``to the vehicle'' case. Pumping frequency 12GHz.}
\end{figure}

\begin{figure}[ht!]
\caption{\label{Theta_Phi_to_18GHz_small} Dependence the of power conversion efficiency coefficient $\mu_S$ on several pumping angles, in the ``to the vehicle'' case. Pumping frequency 18GHz.}  
\end{figure}

\begin{figure}[ht!] 
\caption{\label{Theta_Phi_from_12GHz} Dependence of the power conversion efficiency coefficient $\mu$ on optimal angle. ``From the vehicle''. Pumping frequency 12GHz.}
\end{figure}

\begin{figure}[ht!]
\caption{\label{Theta_Phi_from_18GHz} Dependence of power conversion efficiency coefficient $\mu$ on optimal angle. ``From the vehicle''. Pumping frequency 18GHz.}  
\end{figure}

\begin{figure}[ht!]
\caption{\label{antenna} Schematic plot of beam diffraction.}
\end{figure}

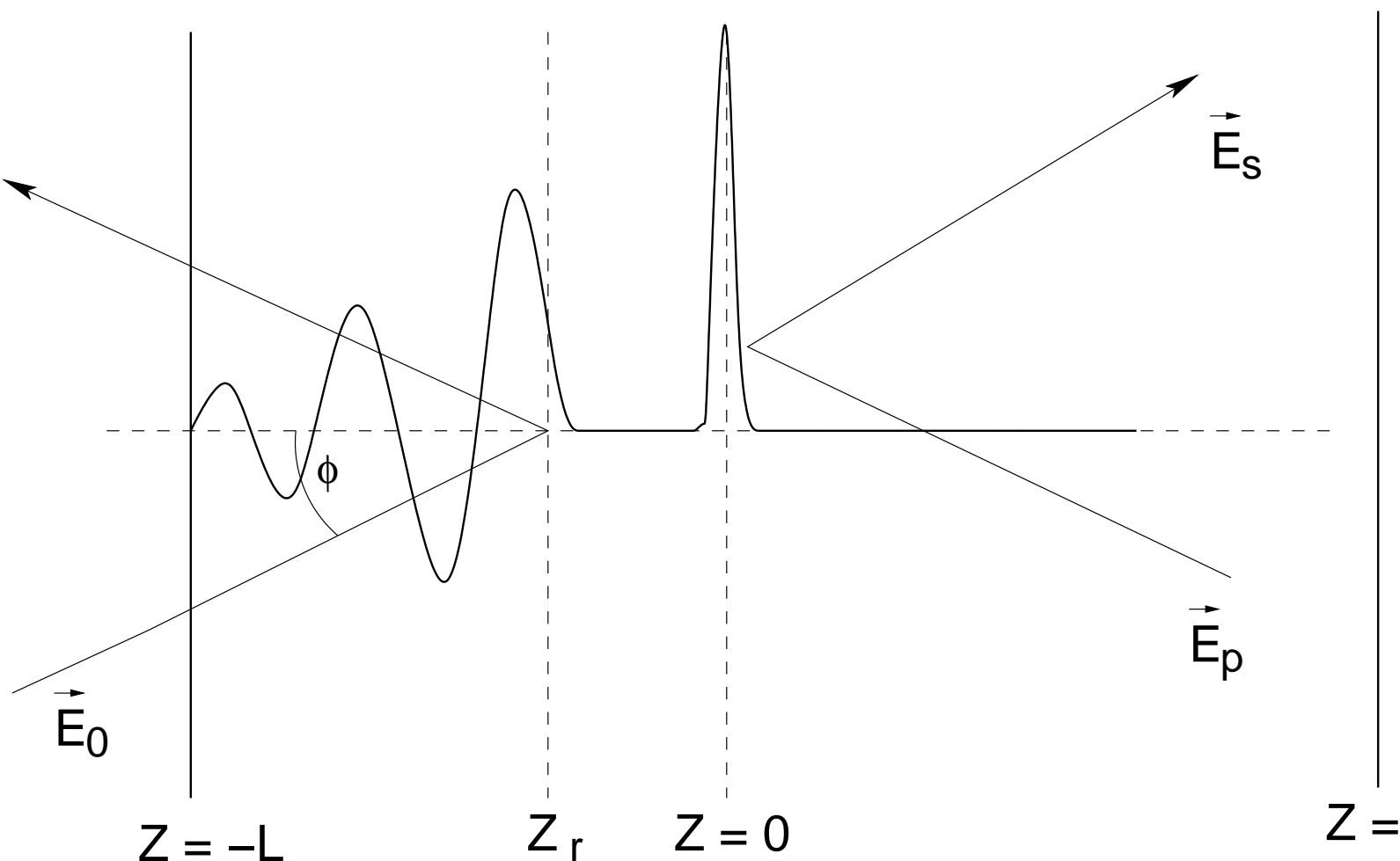
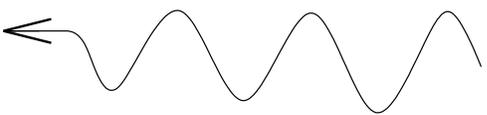
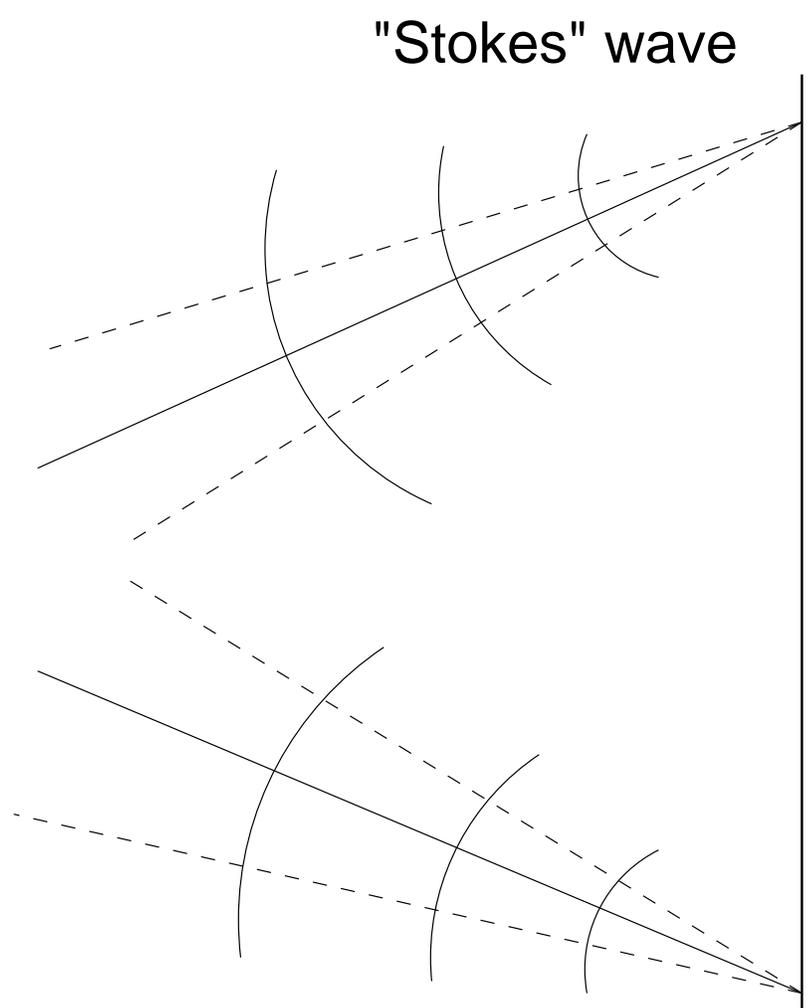
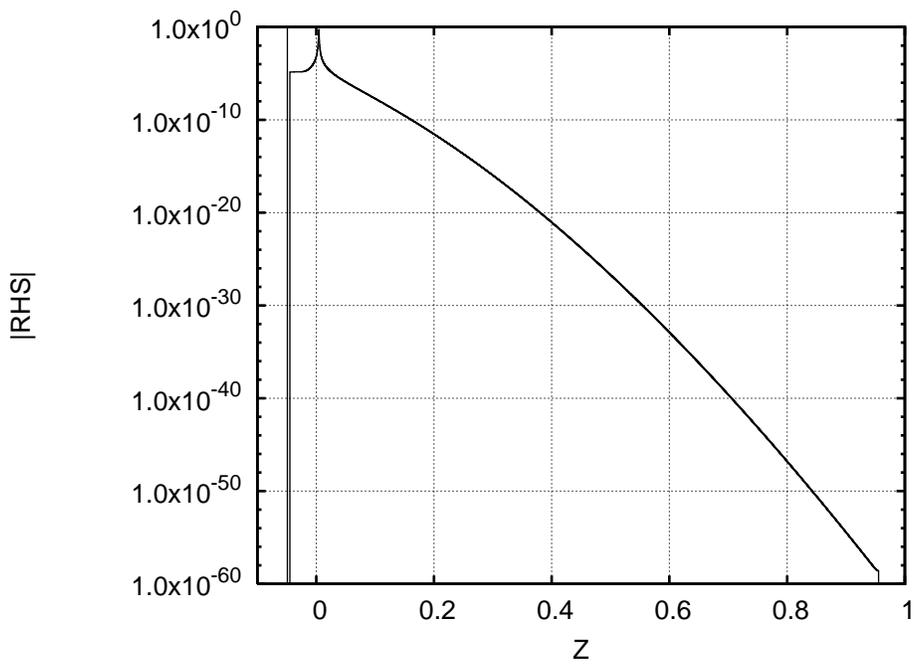
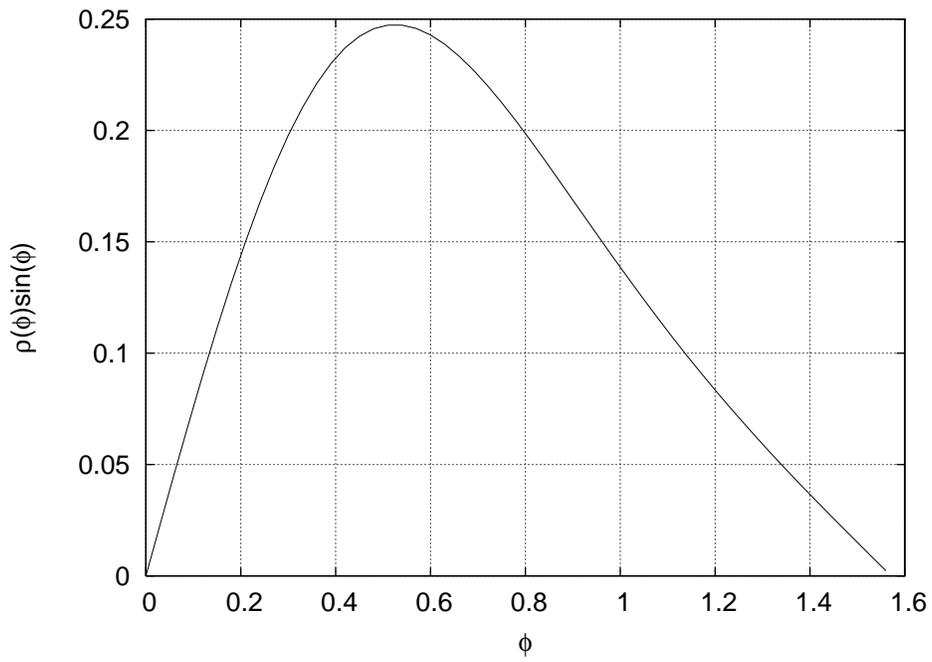
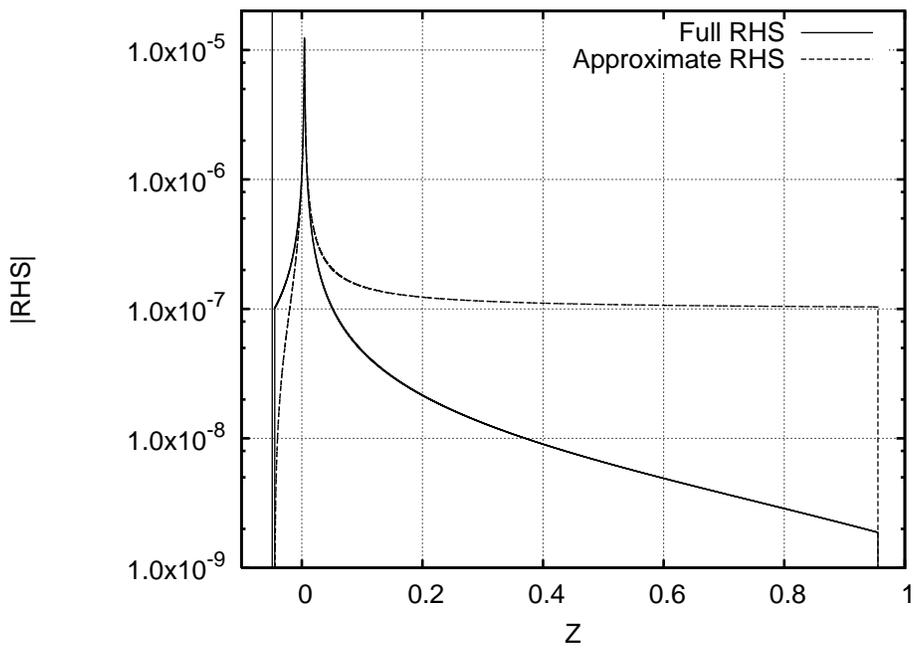
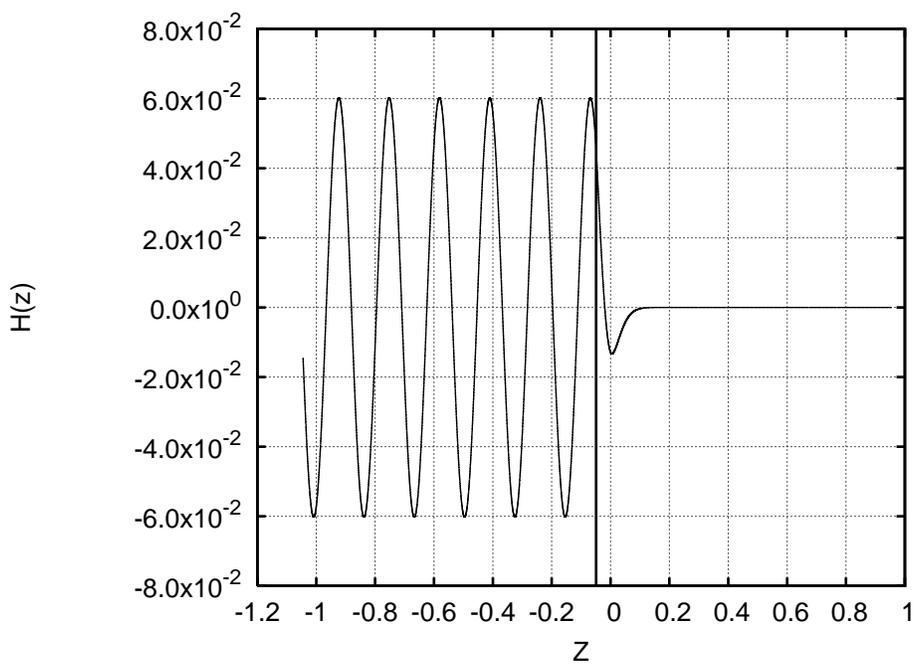
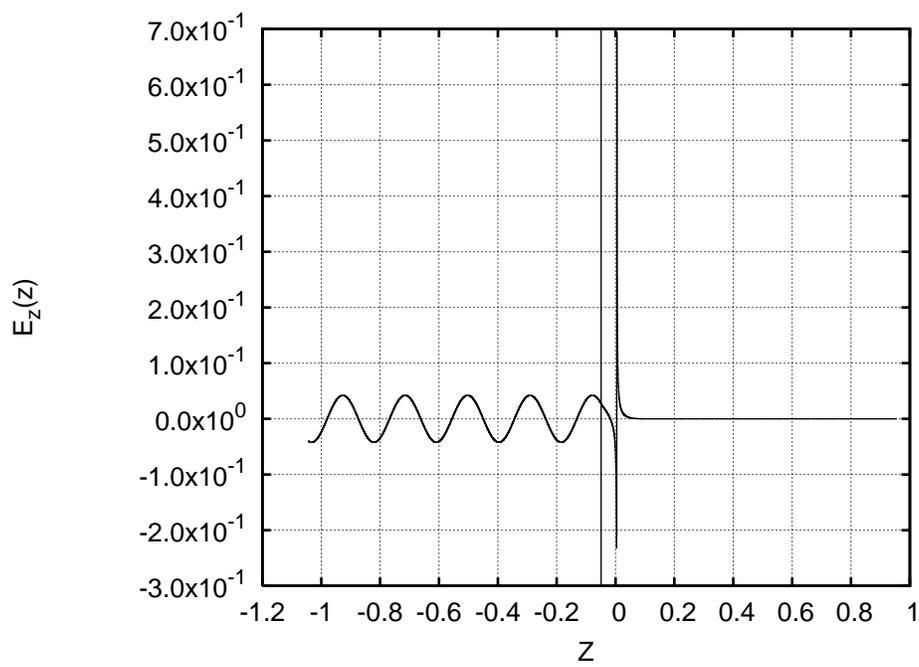
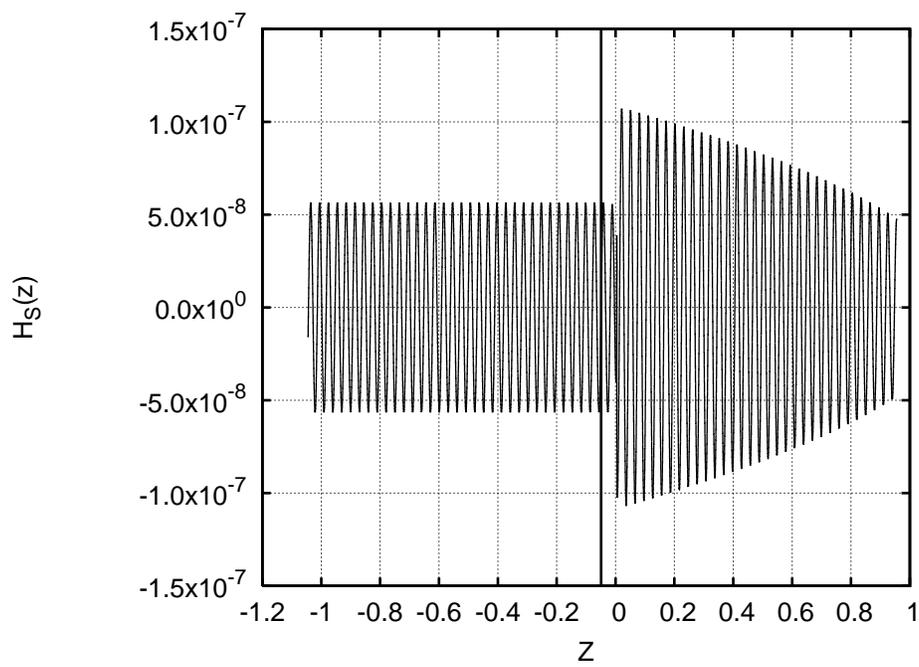
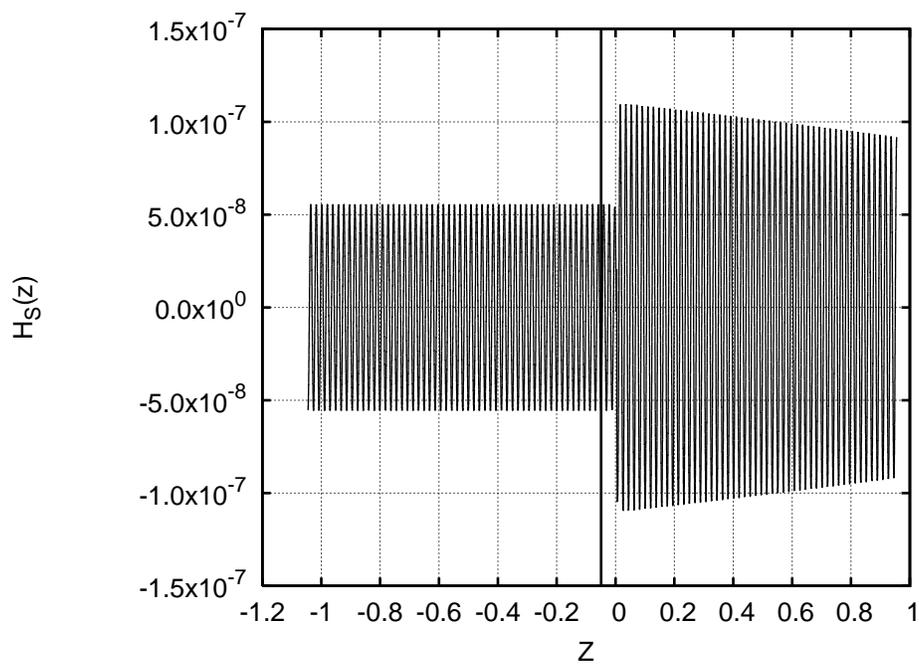
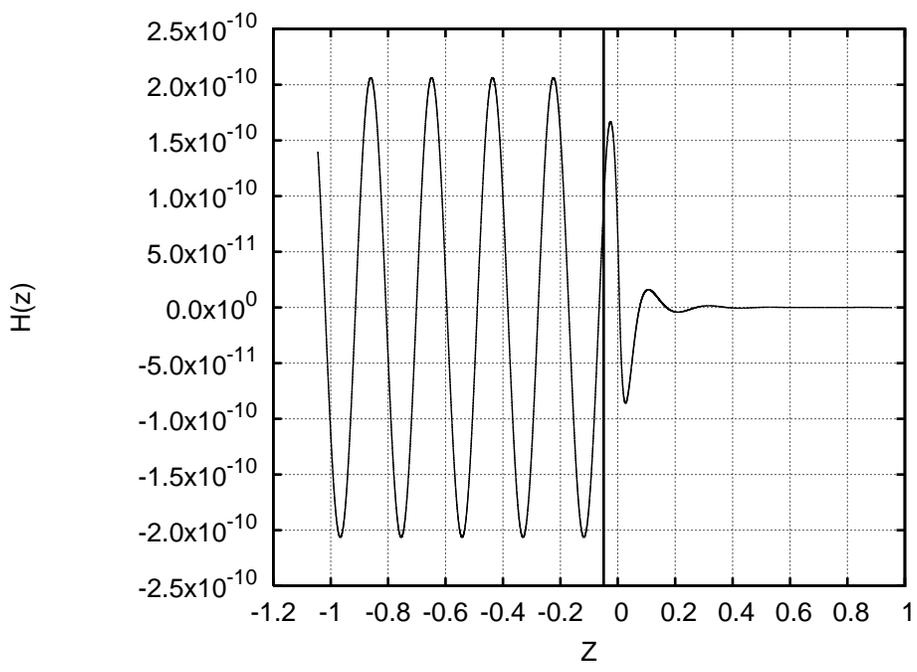
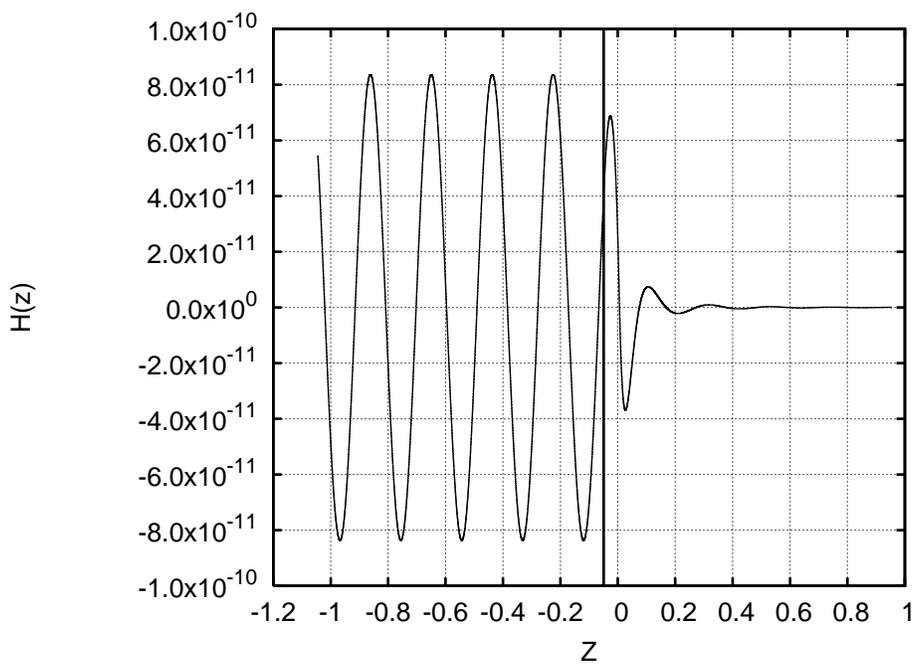
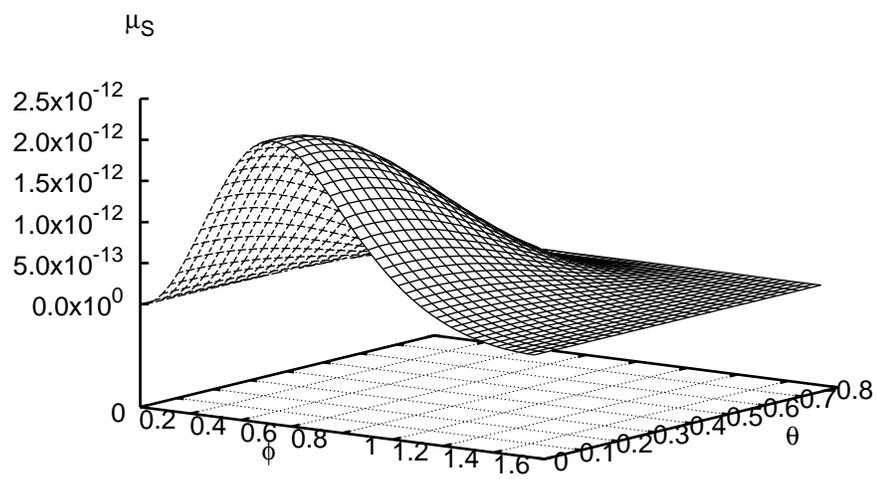
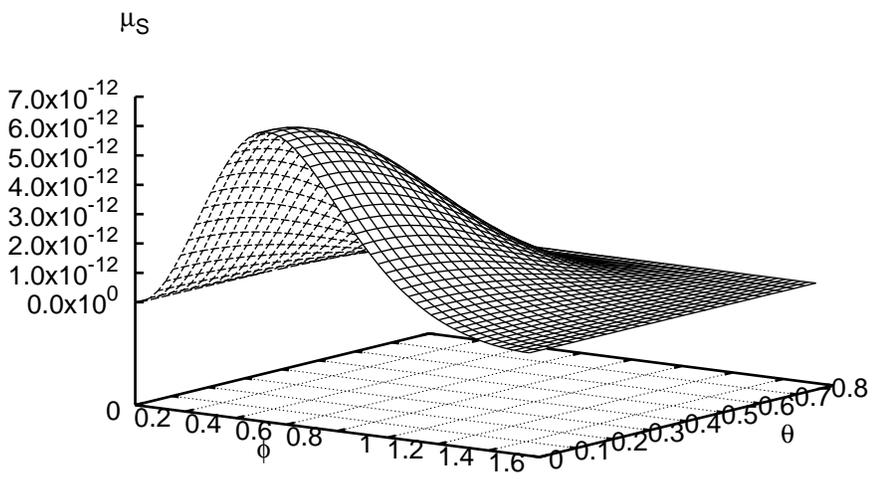
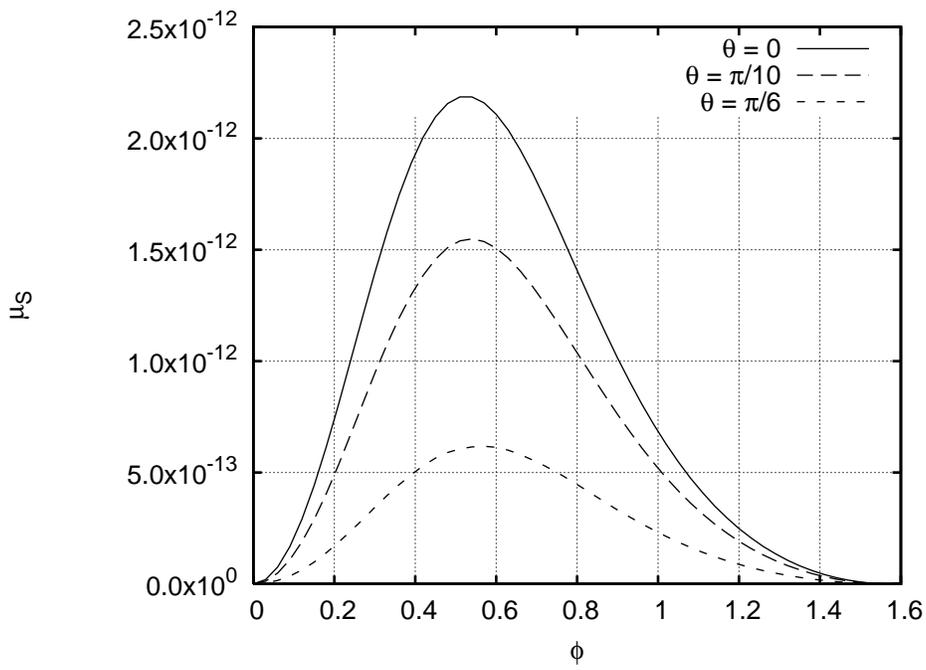
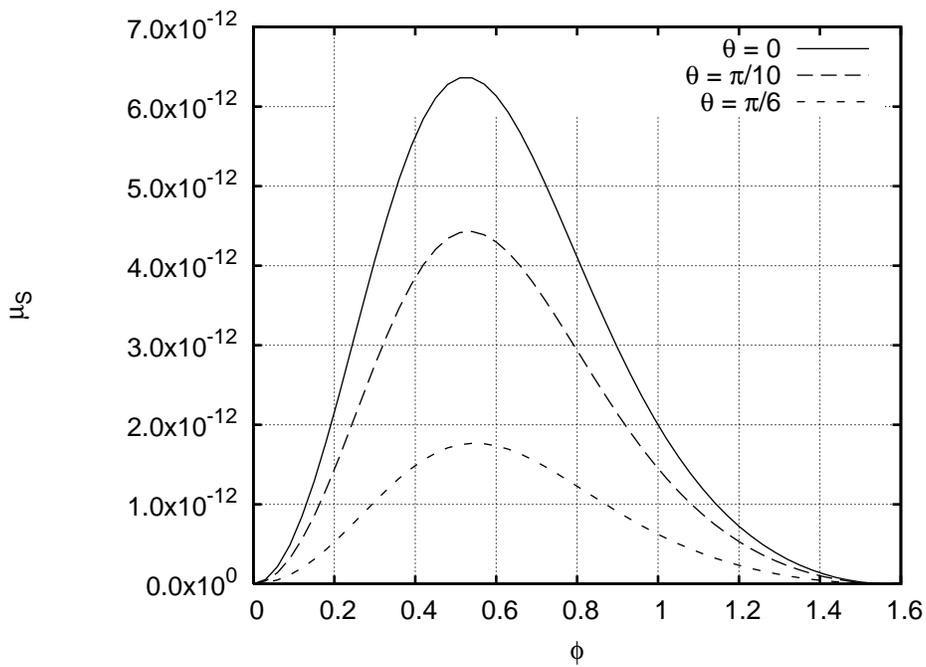
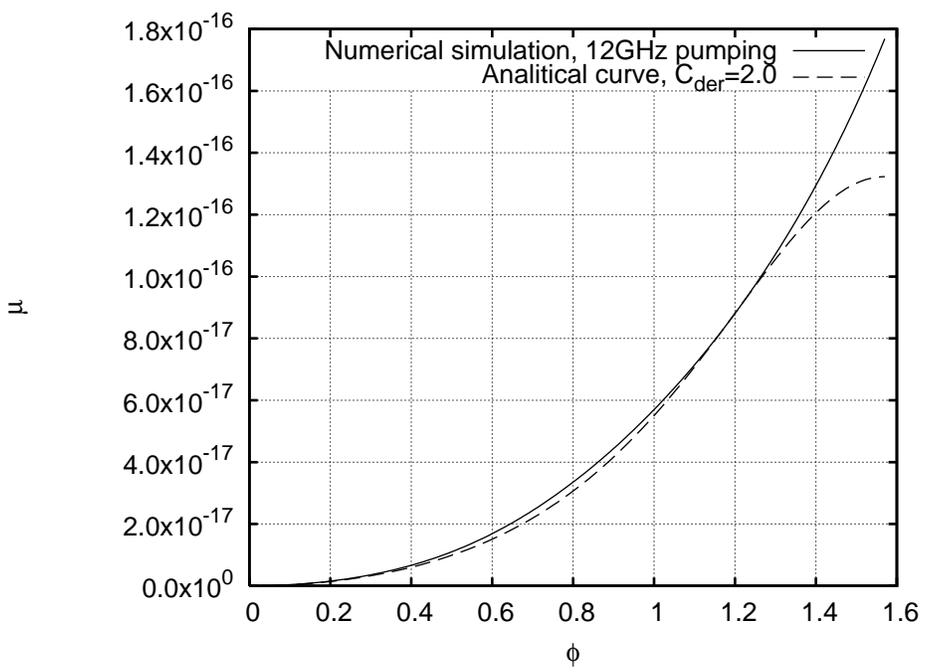
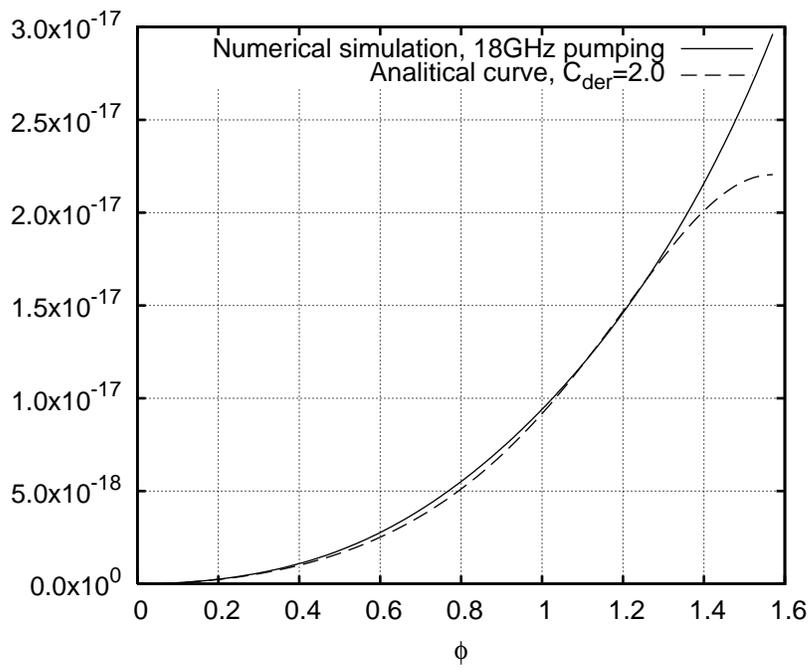
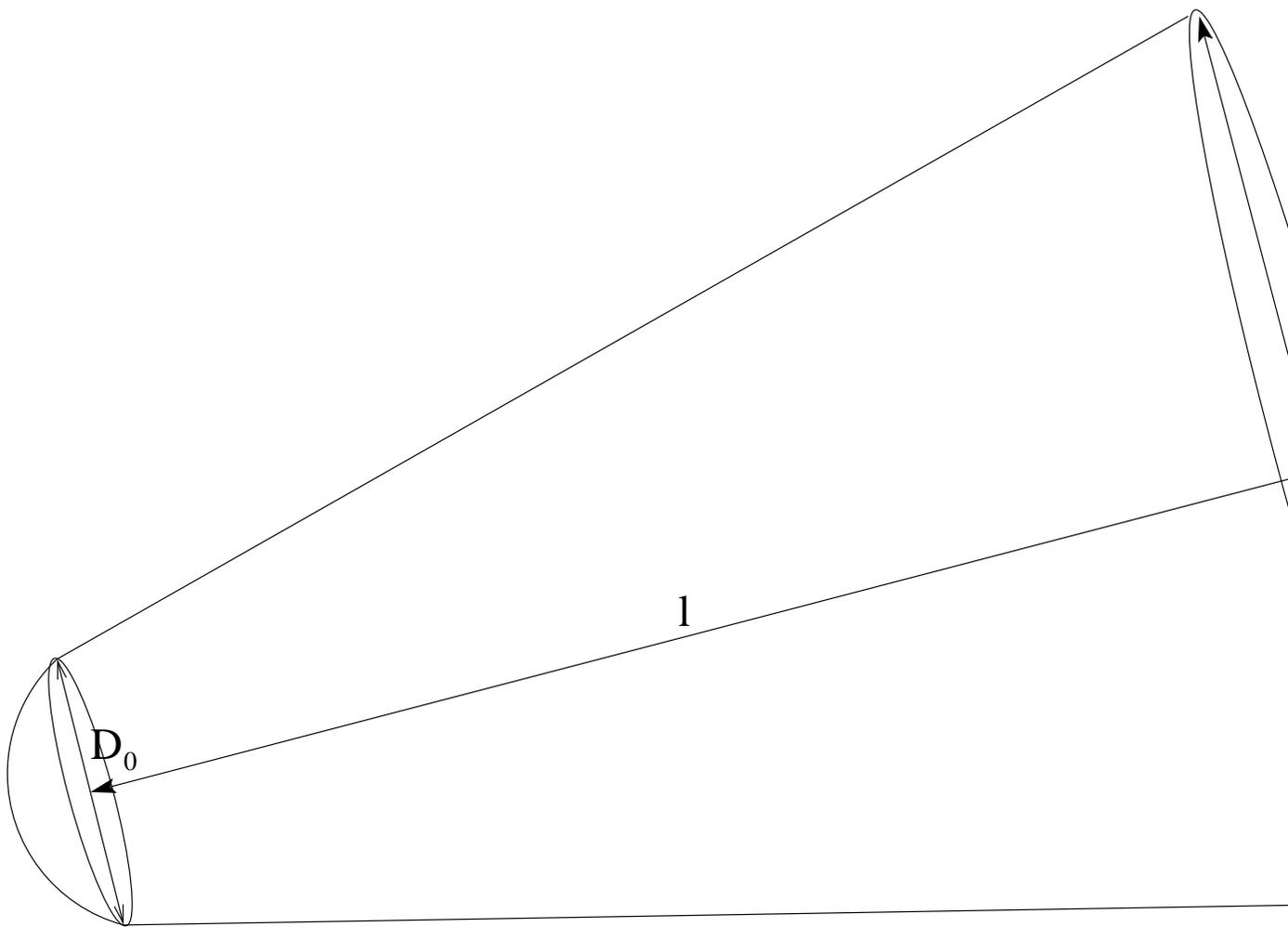
\end{document}